\documentclass[12pt]{article}
\usepackage{cite,epsfig,amssymb} 

\def\ltsim{\lower3pt\hbox{$\, \buildrel < \over \sim \, $}}
\def\gtsim{\lower3pt\hbox{$\, \buildrel > \over \sim \, $}}
\def\be{\begin{equation}}
\def\ee{\end{equation}}
\def\ba{\begin{eqnarray}}
\def\ea{\end{eqnarray}}
\def\de{\partial}
\def\ga{\mathrel{\raise.3ex\hbox{$>$\kern-.75em\lower1ex\hbox{$\sim$}}}}
\def\la{\mathrel{\raise.3ex\hbox{$<$\kern-.75em\lower1ex\hbox{$\sim$}}}}

\newcommand{\chib}{\mbox{\boldmath$\chi$}}

\begin{document}

\baselineskip=16pt   
\begin{titlepage}  
\rightline{IHES/P/02/34}
\rightline{OUTP-02-07P}
\rightline{LPT-Orsay-02/54}
\rightline{hep-th/0206044}  
\rightline{June 2002} 
\begin{center}  
  
\vspace{0.1cm}  
  
\large {\bf Non-linear bigravity and cosmic acceleration}  

\vspace*{2mm}  
\normalsize  
   
{\bf Thibault Damour$^{a,}$\footnote{damour@ihes.fr}, Ian I. Kogan$^{a,b,c,}$\footnote{iankogan@ihes.fr} and  Antonios   Papazoglou$^{a,}$\footnote{antpap@ihes.fr}}

\smallskip   
\medskip   
{\it $^a$Institut des Hautes \'Etudes Scientifiques}\\  
{\it 35 route de Chartres, 91440 Bures-sur-Yvette, France}  

{\it $^b$Theoretical Physics, Department of Physics, Oxford University}\\  
{\it 1 Keble Road, Oxford, OX1 3NP,  UK}  

{\it $^c$Laboratoire de Physique Th\'eorique, Universit\'e de Paris XI}\\  
{\it 91405 Orsay C\'edex, France}  

\vskip0.15in \end{center}  
   
\centerline{\large\bf Abstract}

We explore the cosmological solutions of classes of non-linear bigravity theories. These  theories are defined by effective four-dimensional Lagrangians describing the coupled dynamics of two metric tensors, and containing, in the linearized limit, both a massless graviton and an ultralight one. We focus on two paradigmatic cases: the case where the coupling between the two metrics is given by a Pauli-Fierz-type mass potential, and the case where this coupling derives from five-dimensional brane constructions. We find that cosmological evolutions in bigravity theories can be described in terms of the dynamics of two ``relativistic particles'', moving in a curved Lorenzian space, and connected by some type of nonlinear ``spring''. Classes of bigravity cosmological evolutions exhibit a ``locking'' mechanism under which the two metrics ultimately stabilize in a bi-de-Sitter configuration, with relative (constant) expansion rates. In the absence of matter, we find that a generic feature of bigravity cosmologies is to exhibit a period of cosmic acceleration. This leads us to propose bigravity as a source of a new type of dark energy (``tensor quintessence''), exhibiting specific anisotropic features. Bigravity could also have been the source of primordial inflation.

\vspace*{1mm}   

\end{titlepage}

\section{Introduction}

Recently, there has been some interest in multigravity theories \cite{Kogan:1999wc,Gregory:2000jc,Kogan:2000cv,Kogan:2000xc,Kogan:2000vb,Kogan:2001yr}  where gravity is modified at cosmological scales. These theories involve brane configurations in a higher than four dimensional spacetime where normal four dimensional  gravity on the branes is modified  in the far infrared due to the presence of a massive (but ultralight) graviton component in the low energy theory (see \cite{Kogan:2001ub} for a review and \cite{Papazoglou:2001cc} for a detailed presentation).   The attractive feature of these theories is that they provide an alternative observational window to extra dimensional physics which is testable in current observations \cite{Will:1997bb,Binetruy:2000xv,Uzan:2000mz,Bastero-Gil:2001rv,Choudhury:2002pu}. Most importantly, these modifications of gravity are at such scales from which current observations indicate a dark energy component in our universe \cite{Riess:1998cb,Perlmutter:1998np}. It would be tempting to attribute this to the dynamics of a multigravity system.

The basic idea in constructing multigravity models is to localize gravity at the same time in different places along the extra dimension(s). Once one has a higher dimensional brane configuration which localizes gravity, {\textit{e.g.}} \cite{Randall:1999vf,Gherghetta:2000qi}, the low energy effective theory is governed by a massless graviton field. By multilocalizing gravity \cite{Kogan:2001wp} in a superposition of such configurations, the degeneracy of the massless modes will be lifted and the low energy theory will contain apart of a massless mode, a collection of light massive gravitons. How light these gravitons will be depends on how strong the localization in the single graviton  configuration is. In the particular models that have been examined in the literature \cite{Kogan:1999wc,Gregory:2000jc,Kogan:2000cv,Kogan:2000xc,Kogan:2000vb,Kogan:2001yr}, the localization was exponentially strong  and thus the mass splittings between  the light gravitons and the remaining of the Kaluza-Klein (KK) spectrum was exponentially large. This gave the opportunity to realize models which escaped observational bounds and had  interesting phenomenological implications. An alternative mechanism which has similar effects arises in the case where four dimensional gravity is induced on the brane due to quantum loops of  matter living on the brane \cite{Dvali:2000hr,Dvali:2000xg}. In that case, the old result of Sakharov \cite{Sakharov:pk} (for a review of induced gravity see \cite{Adler:1982ri}) was exploited to render the geometry on a brane, embedded in a flat higher dimensional space, four dimensional.

However, so far the models \cite{Kogan:1999wc,Gregory:2000jc,Kogan:2000cv,Kogan:2000xc,Kogan:2000vb,Kogan:2001yr}  where studied at the linearized level which becomes invalid when speaking about cosmological distance dynamics. One clearly has to go beyond the linear theory, which is the main aim of this paper. Nonlinear bigravity theories were first introduced in the seventies as effective descriptions of a sector of hadronic physics \cite{Isham:gm}. It is argued in a companion paper \cite{DK} that nonlinear bigravity theories can arise in several different (purely gravitational) contexts: multibrane configurations, certain classes of Kaluza-Klein models, some types of non-commutative geometry models, {\textit{etc.}} It is, therefore, important to try to delineate what are the generic predictions of classes of bigravity theories. One of the main conclusions of the present paper is that bigravity naturally gives rise to a late period of cosmic acceleration. Bigravity can then be used as a new theoretical model of dark energy (with specific anisotropic features, in certain cases, that make it phenomenologically distinguishable from quintessence models).

Accelerating solutions were also found in the context of one particular model of brane-induced gravity \cite{Dvali:2000hr,Dvali:2000xg}, and their phenomenological consequences have been explored in detail as a possible theoretical model of dark energy \cite{Deffayet:2000uy,Deffayet:2001pu}. Both models share the common feature of modifications of gravity at large scales. However, our work differs in several ways. We study general classes of four-dimensional effective theories instead of one particular five-dimensional model. We study general classes of solutions of these theories including their stability. We find several different types of accelerating solutions some being isotropic, but many featuring a novel type of anisotropic acceleration.

In the present paper, we  will firstly describe the formalism needed for  discussing a general bigravity system. We will be interested in discussing cosmological solutions for such a system. As illustrative models  we will use two types of potentials coupling the two metrics. One which resembles the Pauli-Fierz mass term in the linearized theory and one which is motivated by a higher dimensional brane construction. For these potentials we will be examining the most simple case, by imposing special symmetries, and without including matter. We introduce a description of the coupled cosmological evolution of the two metrics in terms of a ``mechanical'' model: two ``relativistic particles'' connected by a nonlinear ``spring''.  We will see that there is a generic period of acceleration of one or both of the metrics. For the Pauli-Fierz potential we discover two classes of accelerating solutions. One where anisotropies play a crucial role, and one where the cosmology is isotropic. On the other hand, for the brane potential our symmetry requirements only allow for solutions where anisotropies play an important role. In the final state the relative lapse between the two metrics tends to run away to infinity, when using the above illustrative coupling potentials. This run-away signals the breakdown of our effective theory. We discuss more general classes of potentials which naturally lead to a confinement of the relative lapse within a limited range. Such models lead to an interesting ``locking'' mechanism of the evolution of the two metrics. At the end, we briefly discuss the inclusion of matter in the above systems and propose the multigravity scenario as a candidate for a purely gravitational type of dark energy.
 
In the following we adopt the mostly plus metric signature $(-,+,+,+)$ and use the following definition for the Riemann tensor  $R^K_{~\Lambda MN}=\de_M \Gamma^K_{\Lambda
N}-\de_N \Gamma^K_{\Lambda M}+\Gamma^H_{\Lambda
N}\Gamma^K_{MH}-\Gamma^H_{\Lambda M}\Gamma^K_{NH}$. We use  capital letters to label four dimensional spacetime coordinates and lower case letters to label three dimensional space coordinates. For the coordinate basis we use greek letters $M,N,\Lambda,\dots=0,1,2,3$ and $\mu,\nu,\lambda,\dots=1,2,3$,  and for the vierbien latin ones $A,B,C,\dots=0,1,2,3$ and  $a,b,c,\dots=1,2,3$. The reason for using, somewhat unconventionally, $\mu,\nu,\lambda,\dots$ for spatial indices will be explained below.

\section{General bigravity action}

A bigravity system is by definition the gravitational system of two 
coupled metrics which in linearized approximation is reduced to two 
 spin-2 fields, one of which is massless while the other has a small mass\footnote{Actually the second mass can also be zero and the coupling between the two
 metrics will emerge at the non-linear level. For more details see  \cite{DK}.}.
The non-linear description of such a system can be achieved by considering the sum of the Einstein actions for two independent metrics (on the same manifold) and in addition an ultralocal potential term which couples the two metrics (see \cite{Isham:gm}). The action is invariant only under the group of common spacetime diffeomorphisms. This guarantees the presence of one massless spin-2 excitation in the gravitational spectrum\footnote{Depending on the parameters of the bigravity 
Lagrangian the  massless graviton  can  be made  arbitrarily
 weakly coupled, but it always will be in the spectrum for all finite values
 of these parameters \cite{DK}.}.  Adding the coupling of matter fields $\{\Phi_1\}$ to the first metric  and of fields  $\{\Phi_2\}$ to the second one, we have the action:
\be
{\mathcal S}={\mathcal S}_G+{\mathcal S}_M
\label{genactionsum}
\ee
with
\be
{\mathcal S}_G={1 \over 2\kappa_1}\int \sqrt{-g_1}{\mathcal R}[{\bf g}_1]+{1 \over 2\kappa_2}\int \sqrt{-g_2}{\mathcal R}[{\bf g}_2]-\mu^4 \int(g_1g_2)^{1/4}\bar{V}\label{a1}
\ee
\be
{\mathcal S}_M= \int \sqrt{-g_1}{\mathcal L}_1({\bf g}_1,\{\Phi_1\})+\int \sqrt{-g_2}{\mathcal L}_2({\bf g}_2,\{\Phi_2\})\label{a2}\nonumber
\ee
where $\bar{V}$ is a dimensionless scalar function of the relative metric ${\bf g}_1^{-1}{\bf g}_2$  and $[\kappa_1]=[\kappa_2]=M^{-2}$ while $[\mu]=M$. Hence, the generic bigravity model has three dimensionful parameters, $\kappa_1$,   $\kappa_2$ and $\mu$.  In the following we will factor out from the potential term the quantity $(2 \bar{\kappa})^{-1}$, where $\bar{\kappa} \equiv {1 \over 2}(\kappa_1+\kappa_2)$ denotes the average gravitational coupling, and absorb the resulting parameter of mass dimension 2 in $\bar{V}$, {\textit{i.e.}} we work with:
\be
{\mathcal S}_G={1 \over 2\kappa_1}\int \sqrt{-g_1}{\mathcal R}[{\bf g}_1]+{1 \over 2\kappa_2}\int \sqrt{-g_2}{\mathcal R}[{\bf g}_2]-{1 \over 2\bar{\kappa}} \int(g_1g_2)^{1/4}V\label{genaction}
\ee
where now $[V]=M^2$. We have also the  obvious freedom to add cosmological terms for the two metrics, but these can be absorbed in the potential $V$. In the following, we will consider that the potential does not have these cosmological terms, in order to isolate the physics of the  coupling part of the above potential.

It is convenient to introduce special frames $\omega^A=e^A_M dx^M$ with respect to which both metrics are diagonal, {\textit{i.e.}}:
\ba 
ds_1^2=\sum_{A=0}^3\lambda^{(1)}_{A}(\omega^A)^2\\
ds_2^2=\sum_{A=0}^3\lambda^{(2)}_{A}(\omega^A)^2
\ea
where $\lambda^{(1)}_{0}$ and $\lambda^{(2)}_{0}$ are negative. We are then interested in the relative eigenvalues of the two metrics:
\be
\lambda_A \equiv {\lambda^{(2)}_{A} \over \lambda^{(1)}_{A}}\equiv e^{\mu_A}
\ee

The potential $V$, in this notation, is a function of the $\lambda_A$'s, or equivalently of the $\mu_A$'s (see \cite{DK} for more details). The induced energy-momentum tensor from the coupling term of the two metrics is for each of the two metrics:
\be
T^{MN}={2 \over \sqrt{-g}}{\delta S_m \over \delta g_{MN}}
\ee
with $S_m=-{1 \over 2\bar{\kappa}} \int(g_1g_2)^{1/4}V$.  In the non-coordinate basis $\omega^A$, it has the simple form:
\ba
T^{(1)~A}_{~~~A}=-{1 \over \bar{\kappa}}e^{\sigma_1/4}\left({V(\mu_B) \over 4 }-{\de V(\mu_B) \over \de \mu_A}\right)~~~ ({\rm no~sum})\label{energy1}\\
T^{(2)~A}_{~~~A}=-{1 \over \bar{\kappa}}e^{-\sigma_1/4}\left({V(\mu_B) \over 4 }+{\de V(\mu_B) \over \de \mu_A}\right)~~~ ({\rm no~sum})
\label{energy2}
\ea
where  $\sigma_1 \equiv \sum_A\mu_A$. This gives rise to effective energy densities $\rho_1=-T^{(1)~0}_{~~~0}$, $\rho_2=-T^{(2)~0}_{~~~0}$ and effective pressures $P_1^a=T^{(1)~a}_{~~~a}$, $P_2^a=T^{(2)~a}_{~~~a}$ generated for the two metrics by the coupling potential (and both measured in their respective physical units).

The only restriction we shall impose on the form of $V$ is that at quadratic level it should reproduce the Pauli-Fierz mass term if the two metrics are expanded around a flat background, since this is the only ghost-free Lagrangian for a spin-2 field. Apart from this, the form of $V$ is rather unconstrained. For example, a possible potential term was considered long ago in the ``strong gravity'' theory \cite{Isham:gm}  and reads:
\be
V=m^2(-g_1)^{u-1/4}(-g_2)^{v-1/4}\left\{{\rm tr}[(({\bf g}_2^{-1}-{\bf g}_1^{-1}){\bf g}_1)^2]-({\rm tr}[({\bf g}_2^{-1}-{\bf g}_1^{-1}){\bf g}_1])^2 \right\}
\ee
with $u+v={1 \over 2}$.  We should point out here that this particular choice of potential is, contrary to the potentials we shall consider, asymmetric in the exchange of ${\bf g}_1$ and ${\bf g}_2$.

In the general  case  in four dimensions, the potential depends on  $\sigma_1$,  $\sigma_2$, $\sigma_3$, $\sigma_4$,  {\textit{i.e.}}  $V=V(\sigma_1,\sigma_2,\sigma_3,\sigma_4)$, where $\sigma_n \equiv \sum_A\mu_A^n$. In the following we will focus on the class of potentials which depend only on $\sigma_1$ and $\sigma_2$, {\textit{i.e.}} $V=V(\sigma_1,\sigma_2)$. Note that for these potentials, (\ref{energy1}) yields the following energy density and pressures for the first metric:
\ba
&\rho_1={1 \over \bar{\kappa}}e^{\sigma_1/4}\left({1 \over 4}V-\de_{\sigma_1} V- 2 \mu_0 ~\de_{\sigma_2} V \right)\\
&P_1^a=-{1 \over \bar{\kappa}}e^{\sigma_1/4}\left({1 \over 4}V-\de_{\sigma_1} V- 2 \mu_a ~\de_{\sigma_2} V \right)
\ea
 while  (\ref{energy2}) yields for the second metric:
\ba
&\rho_2={1 \over \bar{\kappa}}e^{-\sigma_1/4}\left({1 \over 4}V+\de_{\sigma_1} V+ 2 \mu_0 ~\de_{\sigma_2} V \right)\\
&P_2^a=-{1 \over \bar{\kappa}}e^{-\sigma_1/4}\left({1 \over 4}V+\de_{\sigma_1} V+ 2 \mu_a ~\de_{\sigma_2} V \right)
\ea

This class of potentials is the minimal class which can reproduce the Pauli-Fierz mass term in the limit where $g^1_{MN}$ and  $g^2_{MN}$ are expanded around the {\textit{same}} flat metric $\eta_{MN}$. The particular combination of $\sigma_1$ and $\sigma_2$ which accomplishes this is:
\be
V(\sigma_1,\sigma_2)={m_{PF}^2 \over 8}(\sigma_2-\sigma_1^2)
\label{PF}
\ee

Indeed, expanding the two metrics around a common flat background as $g^1_{MN}=\eta_{MN}+\sqrt{2\kappa_1}h^1_{MN}$ and $g^2_{MN}=\eta_{MN}+\sqrt{2\kappa_2}h^2_{MN}$, and introducing $x_1=\sqrt{\kappa_1 \over \kappa_1+\kappa_2}$ and $x_2=\sqrt{\kappa_2 \over \kappa_1+\kappa_2}$ satisfying $x_1^2+x_2^2=1$, we find that the combination  $h^0_{MN} \equiv x_2 h^1_{MN}+x_1 h^2_{MN}$ is massless while $h^m_{MN} \equiv x_1 h^1_{MN}-x_2 h^2_{MN}={1 \over \sqrt{2(\kappa_1+\kappa_2)}}(g^1_{MN}-g^2_{MN})$ has a Pauli-Fierz mass term:
\be
-{1 \over 2\bar{\kappa}}(g_1g_2)^{1/4}V=- {m_{PF}^2 \over 4} \left(h^{MN}_m h^m_{MN}-h_m^2\right)
\ee
The above potential is the simplest choice that one could think about and hence we will mainly focus on it in the following. We shall call the above potential the ``Pauli-Fierz potential'' and examine the physics when  the quantities  $\sigma_1$ and $\sigma_2$ are not necessarily small. 

We shall also consider the potential that  arises as the effective four dimensional description of the brane motivated  bigravity scenario \cite{Kogan:1999wc}. This potential has the form \cite{DK}:
\ba
V=m^2\left[\cosh B-\cosh A\right],\label{branepot}\\{\rm with} ~~~A={\sigma_1 \over 4} ~~~ {\rm and} ~~~ B={1 \over 2 \sqrt{3}}\sqrt{\sigma_2-{\sigma_1^2 \over 4}}
\ea
The potential $V$ in this case is a  function solely of $\sigma_1$ and $\sigma_2$. In addition, it is symmetric under the exchange of ${\bf g}_1$ and $\bf{g}_2$ (or equivalently it is an even function of the $\mu_A$'s). For small  $\sigma_1$ and $\sigma_2$ we get exactly the Pauli-Fierz potential (\ref{PF}), with the change of notation:
\be
m \equiv \sqrt{3}~m_{PF}
\ee

The discussion in \cite{DK} of the Arnowitt-Deser-Misner Hamiltonian approach to bigravity has shown that a crucial feature of the potential $V({\bf g}_1^{-1}{\bf g}_2)$ is its ability (or lack of ability) to ``confine'', within a limited range, the variation of the relative lapse $n \equiv (N_2/N_1)^{1/2}$, by means of its (algebraic) equation of motion. We have found that the confining capabilities of both the simple ``Pauli-Fierz'' potential (\ref{PF}) and the ``brane-motivated'' one (\ref{branepot}) are too weak to prevent the relative lapse (appearing as $\gamma=2 \log n$ in our notation below) from running away to infinity in a finite (proper) time in generic solutions. This led us to introduce and study modified versions of the above two potentials. For instance, instead of the purely ``quadratic'' (in the $\mu_A$'s) Pauli-Fierz potential (\ref{PF}), we shall consider a Pauli-Fierz potential augmented by ``quartic'' terms, say (with the notation $m \equiv \sqrt{3}~m_{PF}$):
\be
V(\sigma_1,\sigma_2)={m^2 \over 24}(\sigma_2-\sigma_1^2+\lambda \sigma_2^2)\label{defpot}
\ee
where $\lambda$ is a {\textit{positive}} parameter of order unity.

In the above cases, even though the potentials are symmetric in ${\bf g}_1$ and ${\bf g}_2$, they can couple asymmetrically to matter if   $\kappa_1$ and  $\kappa_2$ differ. For the sake of simplicity in the following we constrain our study in the case where  $\kappa_1=\kappa_2$, where we obtain what is known as symmetric bigravity. Additionally, we normalize $\kappa_1=\kappa_2=1/2$ and leave the only dimensionful parameter of the effective action to be $m \equiv \sqrt{3}~m_{PF}$.

\section{Cosmological ans\"atze}

For the above choices of potential we wish to study the cosmological evolution of the bigravity system. A first assumption that we make is that the two metrics depend only on time. Then the line elements will have the form:
\ba 
ds_1^2=&-e^{2\gamma_1}(d x^0)^2&+~\chi^{1}_{\mu\nu}(dx^{\mu}+b_1^{\mu}dx^0)(dx^{\nu}+b_1^{\nu}dx^0)\\
ds_2^2=&-e^{2\gamma_2}(d x^0)^2&+~\chi^{2}_{\mu\nu}(dx^{\mu}+b_2^{\mu}dx^0)(dx^{\nu}+b_2^{\nu}dx^0)
\ea
where all the functions indicated are only time-dependent. The residual symmetry of such cosmological metrics are the following common diffeomorphisms:
\ba
&x^{0}=&f(x^{0 \prime})\label{dif1}\\
&x^{\mu}=&x^{\mu \prime}+\xi^{\mu}(x^{0 \prime})\label{dif2}
\ea
where $f(x^{0 \prime})$ and $\xi^{\mu}(x^{0 \prime})$ are arbitrary functions of time. Under such common diffeomorphisms the various fields transform as:
\ba
&e^{2 \gamma_1} \to e^{2 \gamma_1}f^{\prime 2}~~,~~e^{2 \gamma_2} \to e^{2 \gamma_2}f^{\prime 2}~~,~~\chi^{1}_{\mu\nu} \to \chi^{1}_{\mu\nu}~~,~~\chi^{2}_{\mu\nu} \to \chi^{2}_{\mu\nu}~~,\nonumber\\
&b_1^{\mu} \to b_1^{\mu}f'+(\xi^{\mu})'~~,~~b_2^{\mu} \to b_2^{\mu}f'+(\xi^{\mu})'\label{tran}
\ea
so that the quantities $\gamma \equiv \gamma_2-\gamma_1$, $\chi^{1}_{\mu\nu}$, $\chi^{2}_{\mu\nu}$ and  $b^{\mu} \equiv e^{-\bar{\gamma}}(b_2^{\mu}-b_1^{\mu})$, where $\bar{\gamma}={\gamma_1+\gamma_2 \over 2}$, are invariant. If we where to consider only the Einstein terms in the action (\ref{a1}) together with the matter terms (\ref{a2}), $b_1^{\mu}$ and $b_2^{\mu}$ would only enter (modulus surface terms) as Lagrange multipliers in front of the momentum constraints. As these constraints identically vanish within the simple (``Bianchi Type I'') cosmological solutions that we consider, we conclude that $b_1^{\mu}$ and $b_2^{\mu}$ only enter the action through the potential $V$. Moreover, because of the transformation properties exhibited in  (\ref{tran}), the invariant $V$ can only be a function of $\gamma \equiv \gamma_2-\gamma_1$, $\chi^{1}_{\mu\nu}$, $\chi^{2}_{\mu\nu}$ and the combination $b^{\mu} \equiv e^{-\bar{\gamma}}(b_2^{\mu}-b_1^{\mu})$ with $\bar{\gamma}={\gamma_1+\gamma_2 \over 2}$ as above. Taking into account the fact that ${\rm det} g^1_{MN}=-e^{2\gamma_1}{\rm det} \chi^1_{\mu \nu}$ (similarly for the second metric), we see that the shift vectors enter the action only through $e^{\bar{\gamma}} ({\rm det} \chi^1_{\mu \nu})^{1/4}( {\rm det} \chi^2_{\mu \nu})^{1/4}V(\gamma,b^{\mu},\chi^1_{\mu \nu},\chi^2_{\mu \nu})$. As $V$ is a scalar, it can involve the vector $b^{\mu}$ only through some scalar combinations such as $h_{\mu \nu}b^{\mu}b^{\nu}$, $k_{\mu \nu \kappa \lambda}b^{\mu}b^{\nu}b^{\kappa}b^{\lambda}$, $\dots$,  where $h_{\mu \nu}$, $k_{\mu \nu \kappa \lambda}$, $\dots$ are made from $\chi^1_{\mu \nu}$, $\chi^2_{\mu \nu}$ and $\gamma$. As $V$ is by assumption a smooth function of ${\bf g}_1^{-1}{\bf g}_2$, it will be a smooth function of the scalars made with $b^{\mu}$, so that we can write:
\be
V(b^{\mu})=V_0+h_{\mu \nu}b^{\mu}b^{\nu}+k_{\mu \nu \kappa \lambda}b^{\mu}b^{\nu}b^{\kappa}b^{\lambda}+{\mathcal{O}}(b^6)\label{bpot}
\ee
where  $V_0$, $h_{\mu \nu}$, $k_{\mu \nu \kappa \lambda}$, depend only on $\chi^1_{\mu \nu}$, $\chi^2_{\mu \nu}$ and $\gamma$. The equation of motion for $b_1^{\mu}$, $b_2^{\mu}$ is simply that $b^{\mu}$ extremize $V(b^{\mu})$, {\textit{i.e.}} ${\de V \over \de b^{\mu}}=0$. We see immediately from (\ref{bpot}) that $b^{\mu}=0$ is always a solution of this equation. In this work, we shall only consider this universal ``perturbative'' solution. Note, however, that, similarly to what happens in Landau's theory of magnetic phase transitions, there might, for some potentials, exist also some ``nonperturbative'' (or ``symmetry breaking'') solutions with $b^{\mu} \neq 0$. In Appendix B we study the equation of motion of $b^{\mu}$ by slightly different approach, and find that ``nonperturbative'' solutions can only exist if the potential $V(\sigma_1,\sigma_2)$ is such that $\de_{\sigma_2}V$ can vanish (which is the case neither for (\ref{PF}) nor for (\ref{branepot})).

Once we know that $b^{\mu}=0$, {\textit{i.e.}} $b_1^{\mu}=b_2^{\mu}$, we can further  use the $\xi_{\mu}(x^{0 \prime})$ gauge freedom to set both $b_1^{\mu}$ and $b_2^{\mu}$ to zero. Then the metrics become:
\ba 
ds_1^2=&-e^{2\gamma_1}(d x^0)^2&+~\chi^{1}_{\mu\nu}dx^{\mu}dx^{\nu}\label{ansatzb1}\\
ds_2^2=&-e^{2\gamma_2}(d x^0)^2&+~\chi^{2}_{\mu\nu}dx^{\mu}dx^{\nu}\label{ansatzb2}
\ea
Note that we still have the gauge freedom $f(x^{0 \prime})$ which changes $\gamma_1$ and $\gamma_2$ but leaves $\gamma_2-\gamma_1$ invariant.

\section{Cosmological dynamics of a bigravity system}

The action describing the dynamics of  $\chi^1_{\mu \nu}$, $\chi^2_{\mu \nu}$ (obtained by inserting the ans\"atze (\ref{ansatzb1}), (\ref{ansatzb2}) into the action (\ref{genaction}); without the matter terms for simplicity) reads ${\mathcal S}=\int d x^0 {\mathcal L}_0$ with: 
\be
{\cal L}_0=e^{-\gamma_1}{\cal K}_1(\chib_1,\dot{\chib}_1)+e^{-\gamma_2}{\cal K}_2(\chib_2,\dot{\chib}_2)-e^{{1 \over 2}(\gamma_1+\gamma_2)}{\cal V}_{12}(\gamma_2-\gamma_1, \chib_1^{-1}\chib_2)\label{klagr}
\ee
where the kinetic terms are (using matrix notation for $\chi^1_{\mu \nu}$, $\chi^2_{\mu \nu}$):
\be
{\cal K}_1(\chib_1,\dot{\chib}_1)={1 \over 4} ({\rm det} \chi^1_{\mu \nu})^{1/2} \left[{\rm tr}(\chib_1^{-1}\dot{\chib}_1)^2-({\rm tr}(\chib_1^{-1}\dot{\chib}_1))^2\right]\label{kterm}
\ee
with a corresponding definition for ${\cal K}_2(\chib_2,\dot{\chib}_2)$ where $\chi^1_{\mu \nu} \to \chi^2_{\mu \nu}$, and:
\be
{\cal V}_{12}(\gamma_2-\gamma_1, \chib_1^{-1}\chib_2)=({\rm det} \chi^1_{\mu \nu})^{1/4}({\rm det} \chi^2_{\mu \nu})^{1/4}V(\gamma_2-\gamma_1, \chib_1^{-1}\chib_2)\label{spring}
\ee
If we replace the two variables $\gamma_1$, $\gamma_2$ by the equivalent combinations $\bar{\gamma}={1 \over 2}(\gamma_1+\gamma_2)$, $\gamma=\gamma_2-\gamma_1$ (so that $\gamma_1=\bar{\gamma}-{1 \over 2}\gamma$,  $\gamma_2=\bar{\gamma}+{1 \over 2}\gamma$), it is easily seen that $\bar{\gamma}$ is a gauge variable whose equation of motion gives the usual zero-energy (Hamiltonian) constraint:
\be
{\mathcal E}=e^{\gamma/2}{\cal K}_1(\chib_1,\dot{\chib}_1)+e^{-\gamma/2}{\cal K}_2(\chib_2,\dot{\chib}_2)+{\cal V}_{12}(\gamma, \chib_1^{-1}\chib_2)=0
\ee
After imposition of this (conserved) zero-energy constraint, we can use the  $f(x^{0 \prime})$ gauge freedom (\ref{dif1}) to set $\bar{\gamma}$ to zero say. By contrast, $\gamma$ is a dynamical variable whose equation of motion is algebraic, because $\gamma$ has no kinetic terms of its own. The equation of motion of $\gamma$ is that ${\cal L}_0(\gamma)$ should be extremized, {\textit{i.e.}} ${\de {\cal L}_0 \over \de \gamma}=0$:
\be
{\de \over \de \gamma}\left[e^{\gamma/2}{\cal K}_1(\chib_1,\dot{\chib}_1)+e^{-\gamma/2}{\cal K}_2(\chib_2,\dot{\chib}_2)-{\cal V}_{12}(\gamma, \chib_1^{-1}\chib_2)\right]=0\label{extrg}
\ee
We shall assume (see below) that the evolution remains in a domain which confines $\gamma$ to a limited range, {\textit{i.e.}} that the shape of the potential $V(\gamma)$ and the values of the kinetic terms  ${\cal K}_1$, ${\cal K}_2$, are such that there exists a bounded solution $\gamma$ of equation (\ref{extrg}) which can be continuously followed during the time evolution. Then the main problem is to discuss the coupled dynamics of the matrices $\chi^1_{\mu \nu}$ and $\chi^2_{\mu \nu}$, obtained (say in the gauge $\bar{\gamma}=0$) from the Lagrangian:
\be 
{\cal L}_0=e^{\gamma/2}{\cal K}_1(\chib_1,\dot{\chib}_1)+e^{-\gamma/2}{\cal K}_2(\chib_2,\dot{\chib}_2)-{\cal V}_{12}(\gamma, \chib_1^{-1}\chib_2)\label{lagrchi}
\ee
For obtaining the equations of motion of $\chi^1_{\mu \nu}$ and $\chi^2_{\mu \nu}$, one can equivalently either treat $\gamma$ as an independent variable, which will be later replaced by the solution of equation (\ref{extrg}), or as a function of  $\chib_1$, $\dot{\chib}_1$, $\chib_2$, $\dot{\chib}_2$ defined by equation (\ref{extrg}). The latter way of viewing $\gamma$ transforms  (\ref{lagrchi}) in a very complicated and very nonlinear Lagrangian ${\cal L}_{\rm reduced}(\chib_1,\dot{\chib}_1,\chib_2,\dot{\chib}_2)$. For simplicity, using our assumption that $\gamma$ remains bounded, {\textit{i.e.}} ${\cal O}(1)$, we shall view $\gamma$ (until it is replaced in the final equations of motion) as a ``given'' function of time.

In this simplified view, the dynamics of  $\chi^1_{\mu \nu}(x^0)$ and $\chi^2_{\mu \nu}(x^0)$ can be interpreted as  a mechanical model of two ``particles'' (with $\gamma$-dependent, {\textit{i.e.}} time-dependent masses), living in a six-dimensional pseudo-Riemannian space, and connected by some type of (time-dependent) nonlinear ``spring''. Indeed, each symmetric $3 \times 3$ matrix $\chi^i_{\mu \nu}$, $i=1,2$ has six independent components, and the kinetic terms (\ref{kterm}) define a certain (curved) Riemannian metric, with signature $(-,+,+,+,+,+)$. The ($\gamma$-dependent) nonlinear ``spring'' is defined by the potential ${\cal V}_{12}(\chib_1^{-1}\chib_2)$ in equation (\ref{spring}). It is interesting to note that, in the case of the brane potential, the potential  ${\cal V}_{12}(\chib_1^{-1}\chib_2)$ happens to be precisely the {\textit{squared geodesic distance}} between $\chib_1$ and $\chib_2$, as defined by the Riemannian metric $d \chi^2={\cal K}(\chib,d \chib)={\cal K}(\chib,\dot{\chib})dt^2$ \cite{DK}. Therefore, in this case, modulo the $\gamma$-dependent modulations, the cosmological evolution system can be elegantly viewed as the problem of two ``particles'' in a pseudo-Riemannian space, connected by a harmonic ``spring'' as ${\cal V}_{12}(\chib_1^{-1}\chib_2) \propto ({\rm distance})^2$. 

In spite of the possibility of such an elegant formulation, the actual dynamics of the coupled metrics $\chi^1_{\mu \nu}$ and $\chi^2_{\mu \nu}$ is extremely complicated. Therefore, as a first cut towards understanding the main qualitative features of this dynamics, we shall henceforth further specialize the class of solutions we consider by focusing on the metrics which can be simultaneously diagonalized. It is then convenient to parametrize the diagonal components of $\chi^1_{\mu \nu}$ as $\chi^1_{\mu \nu}={\rm diag} \left(e^{2\alpha^{\mu}}\right)$, and those of $\chi^2_{\mu \nu}$ as $\chi^2_{\mu \nu}={\rm diag} \left(e^{2\beta^{\mu}}\right)$. In other words, the two metrics read:
\ba 
ds_1^2=&-e^{2\gamma_1}(d x^0)^2&+\sum_{\mu=1}^3 e^{2\alpha^{\mu}}(dx^{\mu})^2\label{ansatz1}\\
ds_2^2=&-e^{2\gamma_2}(d x^0)^2&+\sum_{\mu=1}^3 e^{2\beta^{\mu}}(dx^{\mu})^2
\label{ansatz2}
\ea
 Note that this class of metrics is more general than the ``bi-Friedmann'' metrics where $\alpha^{\mu}$ and $\beta^{\mu}$ would both be taken to be $\left({\alpha \over 3},{\alpha \over 3},{\alpha \over 3}\right)$ and $\left({\beta \over 3},{\beta \over 3},{\beta \over 3}\right)$ where $\alpha \equiv \sum_{\mu}\alpha^{\mu}$ and $\beta \equiv \sum_{\mu}\beta^{\mu}$. The consideration of unequal $\alpha^{\mu}$,  $\beta^{\mu}$, $\mu=1,2,3$, means that we are considering {\textit{anisotropic}} metrics. As we shall see, anisotropies can play a crucial role in the dynamics of ``bi-cosmology''.

For the restricted metrics (\ref{ansatz1}), (\ref{ansatz2}) the Lagrangian (\ref{klagr}) simplifies to: 
\be
{\mathcal L}_0=e^{\alpha-\gamma_1}~{\mathcal G}_{\mu\nu}\dot{\alpha}^{\mu}\dot{\alpha}^{\nu}+e^{\beta-\gamma_2}~{\mathcal G}_{\mu\nu}\dot{\beta}^{\mu}\dot{\beta}^{\nu}-e^{(\alpha+\beta+\gamma_1+\gamma_2)/2}V(\gamma_2-\gamma_1,\beta^{\mu}-\alpha^{\mu})
\label{actnofix}
\ee
where the contractions are made with the internal metric:
\be
{\mathcal G}_{\mu\nu}v^{\mu}v^{\nu}=\sum_{\mu}(v^{\mu})^2 - \left(\sum_{\mu} v^{\mu}\right)^2
\label{intmetric}
\ee
Note that the inverse metric reads:
\be
{\mathcal G}^{\mu\nu}v_{\mu}v_{\nu}=\sum_{\mu}(v_{\mu})^2-{1 \over 2}\left(\sum_{\mu} v_{\mu}\right)^2
\ee
where the factor $1/2$ would be $1/(D-2)$ in spacetime dimension $D$. We note here that the above metric has a Lorentzian signature $(-,+,+)$ and thus we can think of $\alpha^{\mu}$ and $\beta^{\mu}$ as the worldlines of ``particles'' in a three-dimensional spacetime (instead of the above six-dimensional spacetime).The use of greek indices $\mu=1,2,3$ for labeling the three independent scale factors $a^{\mu}=e^{\alpha^{\mu}}$ was chosen to emphasize the Lorenzian nature of the metric ${\mathcal G}_{\mu\nu}d \alpha^{\mu} d \alpha^{\nu}$ underlying the kinetic terms in (\ref{actnofix}). Actually, the ``Poincar\'e'' symmetry in field space (under transformations preserving the scalar product defined by ${\mathcal G}_{\mu\nu}$) of the Lagrangian (\ref{actnofix}) is broken by two sets of terms. Firstly, the separate appearance of $\alpha \equiv \sum_{\mu}\alpha^{\mu}$ and $\beta \equiv \sum_{\mu}\beta^{\mu}$ in the kinetic terms introduces a preferred covector $n_{\mu}=(1,1,1)$ such that $\alpha=n_{\mu}\alpha^{\mu}$ and $\beta=n_{\mu}\beta^{\mu}$ \footnote{Note that the notations $\alpha=n_{\mu}\alpha^{\mu}$, $\beta=n_{\mu}\beta^{\mu}$ and later  $\sigma=n_{\mu}\sigma^{\mu}$, $\delta=n_{\mu}\delta^{\mu}$ correspond to some linear projection of the ``spacetime'' vectors on a certain timelike direction. They could have been denoted $\alpha_0$, $\beta_0$, {\textit{etc.}}. They should not be confused with the magnitude of the corresponding vectors ${\mathcal G}_{\mu\nu}\alpha^{\mu}\alpha^{\nu}$ which we shall never encounter below.}. This covector is timelike because $n^2={\mathcal G}^{\mu\nu}n_{\mu}n_{\nu}=-3/2$. Actually, we see that, in conformity with the metric $d \chi^2={\cal K}(\chib,d\chib)$ introduced above in the space of metrics $\chi_{\mu \nu}$, the Lagrangian (\ref{actnofix}) features the {\textit{curved}} metric $e^{\alpha}{\mathcal G}_{\mu\nu}d \alpha^{\mu} d \alpha^{\nu}$. We found, however, convenient to think in terms of the (conformally related) {\textit{flat}} metric ${\mathcal G}_{\mu\nu}$. Secondly, the potential $V(\sigma_1,\sigma_2)$ depends on:
\ba
&\sigma_1=&2(\gamma_2-\gamma_1+\beta-\alpha)\\
&\sigma_2=&4(\gamma_2-\gamma_1)^2+4\sum_{\mu}(\beta^{\mu}-\alpha^{\mu})^2
\ea
which also breaks the field space ``Poincar\'e'' symmetry. It is crucial to note that, as indicated in (\ref{actnofix}), $V$ depends only on the differences $\gamma_2-\gamma_1$ and $\beta^{\mu}-\alpha^{\mu}$. This suggests to introduce the sums and differences of the above particle coordinates to be:
\ba
\sigma^{\mu}&=\beta^{\mu}+\alpha^{\mu}\\
\delta^{\mu}&=\beta^{\mu}-\alpha^{\mu}
\ea  

Using the $f(x^{0 \prime})$ diffeomorphisms (\ref{dif1}), we can as said above fix the gauge ambiguity by imposing $\bar{\gamma} \equiv {1 \over 2} (\gamma_1+\gamma_2)=0$, {\textit{i.e.}} by setting $\gamma_2=-\gamma_1=\gamma/2$. In this gauge we are using as time parameter the ``average proper time'' $t$, with $dt=e^{\bar{\gamma}}dx^0$. Then after further defining $\sigma \equiv \sum_{\mu}\sigma^{\mu}$, $\delta \equiv \sum_{\mu}\delta^{\mu}$, we obtain the action  ${\mathcal S}=\int dt {\mathcal L}_t$  with:
\ba
{\mathcal L}_t&=&e^{\sigma/2}\left[ e^{(\gamma-\delta)/2}~\dot{\alpha}^{\mu}\dot{\alpha}_{\mu}+e^{-(\gamma-\delta)/2}~\dot{\beta}^{\mu}\dot{\beta}_{\mu}-V(\gamma,\beta^{\mu}-\alpha^{\mu})\right]\nonumber\\
&=&e^{\sigma/2}\left[ {1 \over 2}\cosh\left({\gamma-\delta \over 2}\right)(\dot{\sigma}^{\mu}\dot{\sigma}_{\mu}+\dot{\delta}^{\mu}\dot{\delta}_{\mu})-\sinh\left({\gamma-\delta \over 2}\right)\dot{\sigma}^{\mu}\dot{\delta}_{\mu}-V(\gamma,\delta^{\mu})\right]
\label{actfix}
\ea
where the contractions are made with the internal metric (\ref{intmetric}). When using such a gauge fixed Lagrangian one must remember to impose separately the constraint coming from the variation of the average lapse function $\bar{\gamma}\equiv {1 \over 2}(\gamma_1+\gamma_2)$. As said above, this constraint is the zero-energy condition, {\textit{i.e.}} ${\mathcal E}=K+V=0$ where $K$ is the total kinetic energy. 
As said above, among the equations of motion, two play a special role in that they do not involve second order derivatives. These are the equations of motion obtained by varying the two lapse functions $\gamma_1$ and $\gamma_2$ (which are essentially the Friedmann equations for the two metrics). They read (in the $\bar{\gamma}=0$ gauge):
\ba
\dot{\alpha}^{\mu}\dot{\alpha}_{\mu}=&-\left({V \over 2}-{\de V \over \de \gamma}\right)e^{-(\gamma-\delta)/2}\\
\dot{\beta}^{\mu}\dot{\beta}_{\mu}=&-\left({V \over 2}+{\de V \over \de \gamma}\right)e^{(\gamma-\delta)/2}
\ea
which, using the definitions (\ref{energy1}), (\ref{energy2}),  is equivalent to the more suggestive expressions:
\ba
e^{\gamma}\dot{\alpha}^{\mu}\dot{\alpha}_{\mu}+\rho_1=0 ~~~ \Rightarrow {d \alpha^{\mu} \over dt_1}{d \alpha_{\mu} \over dt_1}+\rho_1=0 \label{constr1a}\\
e^{-\gamma}\dot{\beta}^{\mu}\dot{\beta}_{\mu}+\rho_2=0 ~~~ \Rightarrow {d \beta^{\mu} \over dt_2}{d \beta_{\mu} \over dt_2}+\rho_2=0\label{constr2a}
\ea
where $dt_1=e^{-\gamma/2}dt=e^{\gamma_1}d x^0$ and  $dt_2=e^{\gamma/2}dt=e^{\gamma_2}d x^0$ are the separate proper time coordinates for the two metrics.

Alternatively, working in terms of the combinations $\bar{\gamma} \equiv {1 \over 2}(\gamma_1+\gamma_2)$ and $\gamma \equiv \gamma_2 -\gamma_1$, the equations of motion obtained by varying respectively $\bar{\gamma}$ and $\gamma$ read:
\ba
e^{-\sigma/2}{\cal E}={1 \over 2}\cosh\left({\gamma-\delta \over 2}\right)(\dot{\sigma}^{\mu}\dot{\sigma}_{\mu}+\dot{\delta}^{\mu}\dot{\delta}_{\mu})-\sinh\left({\gamma-\delta \over 2}\right)\dot{\sigma}^{\mu}\dot{\delta}_{\mu}+V(\gamma,\delta^{\mu})=0\label{constr1b}\\
e^{-\sigma/2}{\de {\cal L}_t \over \de \gamma}={1 \over 2}\sinh\left({\gamma-\delta \over 2}\right)(\dot{\sigma}^{\mu}\dot{\sigma}_{\mu}+\dot{\delta}^{\mu}\dot{\delta}_{\mu})-\cosh\left({\gamma-\delta \over 2}\right)\dot{\sigma}^{\mu}\dot{\delta}_{\mu}-{\de V(\gamma,\delta^{\mu}) \over \de \gamma}=0\label{constr2b}
\ea

The above two sets of equations, (\ref{constr1a}), (\ref{constr2a}) versus (\ref{constr1b}), (\ref{constr2b}),  can be easily checked to be totally equivalent. Note that while equation (\ref{constr1b}) is really a first class constraint, which is preserved by the evolution equations of motion, and only needs to be enforced at some initial time, the equation (\ref{constr2b}) is an algebraic equation which determines $\gamma$ as a function of $\alpha^{\mu}$,  $\dot{\alpha}^{\mu}$,  $\beta^{\mu}$,  $\dot{\beta}^{\mu}$ or equivalently $\sigma^{\mu}$,  $\dot{\sigma}^{\mu}$,  $\delta^{\mu}$,  $\dot{\delta}^{\mu}$. As said above, unless explicitly mentioned otherwise (see below), we assume that we are in a regime where $\gamma$ can be continuously solved in terms of $\sigma^{\mu}$,  $\dot{\sigma}^{\mu}$,  $\delta^{\mu}$,  $\dot{\delta}^{\mu}$. Then the main evolution system is obtained by varying $\alpha^{\mu}$, $\beta^{\mu}$, or equivalently  $\sigma^{\mu}$, $\delta^{\mu}$, in the Lagrangian (\ref{actfix}). See Appendix A for the explicit form of these evolution equations.

\begin{figure}[t]

\begin{center}
\epsfig{file=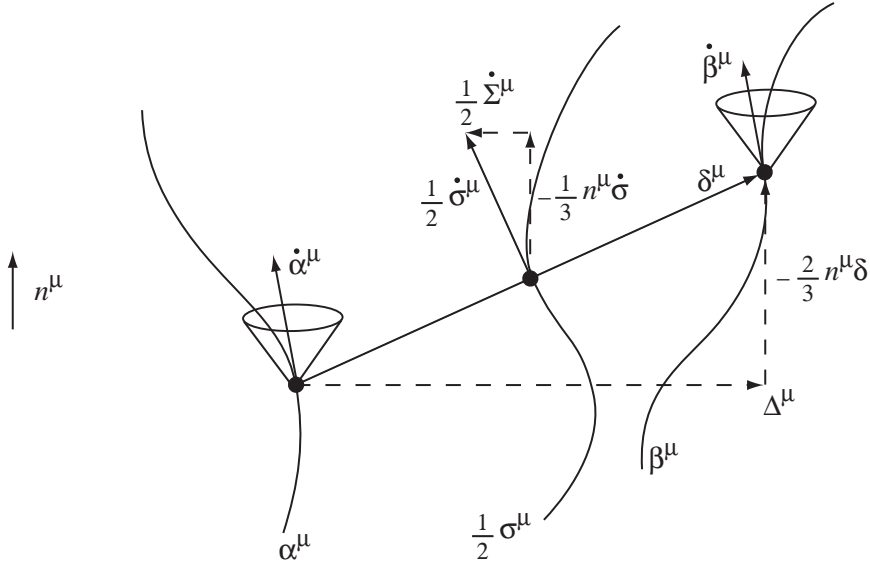}

\caption{General motion of the two ``particle'' worldlines $\alpha^{\mu}$ and $\beta^{\mu}$ in the internal Lorentzian field space. The middle worldline is the ``center of mass'' of the system $\sigma^{\mu}/2$. The decomposition of the ``center of mass'' velocity $\dot{\sigma}^{\mu}/2$ and of the worldline separation $\delta^{\mu}$ in timelike and spacelike components is shown. These projections are performed with respect to the special timelike vector $n^{\mu}$.
\label{genplot}}
\end{center}
\end{figure}

As we said above, the Lagrangian involves, both in the kinetic terms and in the potential, a special covector $n_{\mu}=(1,1,1)$. This makes it useful to decompose the above functions into timelike and spacelike components as:
\ba
\delta^{\mu}&=-{2 \over 3}n^{\mu}\delta+\Delta^{\mu}\label{difference}\\
\sigma^{\mu}&=-{2 \over 3}n^{\mu}\sigma+\Sigma^{\mu}\label{sum}
\ea  
with the property that $\Sigma^{\mu}n_{\mu}=0 \Rightarrow \Sigma^1+\Sigma^2+\Sigma^3=0$  and  $\Delta^{\mu}n_{\mu}=0 \Rightarrow \Delta^1+\Delta^2+\Delta^3=0$. In the definitions (\ref{difference}), (\ref{sum}) there appear the contravariant components of $n_{\mu}$, {\textit{i.e.}} $n^{\mu}={\mathcal G}^{\mu\nu}n_{\nu}=-{1 \over 2}(1,1,1)$ and we recall that  $n_{\mu}n^{\mu}=-{3 \over 2}$. We can easily check that the longitudinal components $\sigma$ and $\delta$ appearing in  (\ref{difference}), (\ref{sum}) are simply $\sigma=\sigma^{\mu}n_{\mu}$ and $\delta=\delta^{\mu}n_{\mu}$. It is helpful to keep in mind the geometrical configuration of Fig.\ref{genplot} which displays the various bimetrical parameters which are relevant for defining the dynamics of bigravity. The elements of this geometrical configuration which play a dominant role in our discussion are: (i) the timelike versus spacelike character of the ``velocity vectors'' $\dot{\alpha}^{\mu}$, $\dot{\beta}^{\mu}$ (with respect to the light cone defined by ${\mathcal G}_{\mu\nu}$), and (ii) the timelike versus spacelike character of the ``separation vector''  $\delta^{\mu} \equiv \beta^{\mu}-\alpha^{\mu}$.

We can define now the fields:
\ba
\Delta_+&=&{3 \over 2}(\Delta^2+\Delta^1)\\
\Delta_-&=&{\sqrt{3} \over 2}(\Delta^2-\Delta^1)
\ea
and respectively for $\Sigma_+$, $\Sigma_-$. Then the Lagrangian can be expressed  as a function of the timelike fields $\sigma$, $\delta$ and the pairs of spacelike ones $\Delta_{\pm}$, $\Sigma_{\pm}$ as:
\ba
{\mathcal L}_{t}=e^{\sigma/2}\left[{1 \over 3}\cosh\left({\gamma-\delta \over 2}\right)(-\dot{\sigma}^{2}-\dot{\delta}^{2}+\dot{\Delta}_+^{2}+\dot{\Delta}_-^{2}+\dot{\Sigma}_+^{2}+\dot{\Sigma}_-^{2})\phantom{aaaaaa}\right.\nonumber\\\left.-{2 \over 3}\sinh\left({\gamma-\delta \over 2}\right)(-\dot{\sigma}\dot{\delta}+\dot{\Delta}_+\dot{\Sigma}_++\dot{\Delta}_-\dot{\Sigma}_-)-V(\gamma,\delta,\Delta_+,\Delta_-)\right]
\ea

As indicated, a generic potential can only be a  function of $\gamma$ and the $\delta^{\mu}$'s, {\textit{i.e.}} of $\gamma$, $\delta$, $\Delta_+$ and $\Delta_-$, hence we see that $\Sigma_+$ and $\Sigma_-$, which correspond to the ``spatial'' location of the center of mass of the two particles $\alpha^{\mu}$, $\beta^{\mu}$, are ignorable coordinates. In fact, in the case of our class of potentials $V(\sigma_1,\sigma_2)$, the dependence of $V$ is even more restricted because:
\ba
&\sigma_1=&2(\gamma+\delta)\label{s1+v}\\
&\sigma_2=&4\left(\gamma^2+{1 \over 3}\delta^2+{2 \over 3}r^2\right)\label{s2+v}
\ea
where $r^2 \equiv {3 \over 2}\Delta^{\mu}\Delta_{\mu} \equiv \Delta_+^2+\Delta_-^2$, so that the potential depends only on the magnitude of the ``spatial distance'' separating the two particles. It is convenient to go to the Routhian formalism and replace the  $\Sigma_+$ and $\Sigma_-$ by their conserved canonical momenta. The Routh functional is then:
\ba
-{\mathcal R}_{t}&=&-\dot{\Sigma}_+ p_{\Sigma_+} - \dot{\Sigma}_- p_{\Sigma_-} + {\mathcal L}_{t}\nonumber\\
&=&e^{\sigma/2}\left\{-{3~(p_{\Sigma_+}^2+p_{\Sigma_-}^2) \over 4\cosh\left({\gamma-\delta \over 2}\right)}e^{-\sigma}-\tanh\left({\gamma-\delta \over 2}\right)(p_{\Sigma_+}\dot{\Delta}_+ + p_{\Sigma_-}\dot{\Delta}_-)e^{-\sigma/2}\right.\nonumber\\
&~&\left.+{1 \over 3}\cosh\left({\gamma-\delta \over 2}\right)\left[{\dot{\Delta}_+^{2}+\dot{\Delta}_-^{2} \over \cosh^2\left({\gamma-\delta \over 2}\right)}-\dot{\sigma}^{2}-\dot{\delta}^{2}\right]+{2 \over 3}\sinh\left({\gamma-\delta \over 2}\right)\dot{\sigma}\dot{\delta}-V(\gamma,\delta,r^2)\right\}~~~~~~~~
\ea
where the conserved momenta are:
\be
p_{\Sigma_\pm}={\de {\mathcal L}_t \over \de \dot{\Sigma}_{\pm} }={2 \over 3}e^{\sigma/2}\left[\cosh\left({\gamma-\delta \over 2}\right)\dot{\Sigma}_{\pm}-\sinh\left({\gamma-\delta \over 2}\right)\dot{\Delta}_{\pm}\right]
\ee

For simplicity, we shall focus on the case where $p_{\Sigma_+}=p_{\Sigma_-}=0$, {\textit{i.e.}} when  the Routhian has no terms linear to $\Delta_+$,  $\Delta_-$. This is the special case where the ``total momentum'' of the two-particle system is directed along the ``vertical'' time axis $n^{\mu}$. As the formal ``Poincar\'e'' invariance of the dynamics is broken, note that going to such  a ``center of mass'' frame is a real restriction on the class of solutions that we examine. Then there exists another conserved quantity because of the rotational symmetry of the system with respect to  $n_{\mu}$. This can be seen by defining:
\be
\Delta_+=r \sin\theta ~~~,~~~  \Delta_- = r \cos\theta
\ee
Then the Routhian is written as:
\be
-{\mathcal R}_{t}=e^{\sigma/2}\left\{{1 \over 3}\cosh \left({\gamma-\delta \over 2}\right) \left[{\dot{r}^{2}+r^2 \dot{\theta}^{2} \over \cosh^2\left({\gamma-\delta \over 2}\right)}-\dot{\sigma}^{2}-\dot{\delta}^{2}\right]+{2 \over 3}\sinh\left({\gamma-\delta \over 2}\right)\dot{\sigma}\dot{\delta}-V(\gamma,\delta,r^2)\right\}
\ee
The potential depends only in $r$, thus there exists a conserved ``angular momentum'' $p_{\theta}$ given by:
\be
p_{\theta}=-{\de {\mathcal R}_{\tau} \over \de \dot{\theta }}={2 \over 3} {r^2 \dot{\theta} \over \cosh\left({\gamma-\delta \over 2}\right)}e^{\sigma/2}
\ee
In analogy with our previous treatment, we can define a new Routhian eliminating the new ignorable coordinate: 
\ba
-\bar{{\mathcal R}}_{t}&=&-\dot{\theta}p_{\theta}-{\mathcal R}_{t}\nonumber\\
&=&e^{\sigma/2}\left\{-{3 p_{\theta}^2\over 4 r^2}\cosh \left({\gamma-\delta \over 2}\right)e^{-\sigma}+{1 \over 3}\cosh \left({\gamma-\delta \over 2}\right) \left[{\dot{r}^{2} \over \cosh^2\left({\gamma-\delta \over 2}\right)}-\dot{\sigma}^{2}-\dot{\delta}^{2}\right]\right.\nonumber\\&~&\left.~~~~~~~+{2 \over 3}\sinh\left({\gamma-\delta \over 2}\right)\dot{\sigma}\dot{\delta}-V(\gamma,\delta,r^2)\right\}
\ea

In the most simple case we can have also $p_{\theta}=0$, where the motion of the two particles is planar. In this case the variable $r$ (which represents say, $\Delta_-$, if $\theta={\rm const.}=0$) can be considered as varying on the full real line, from $-\infty$ to $+\infty$. We will mainly discuss this particular case and from now on, we will indicate by ${\mathcal L}$ the simplified Routhian:
\be
{\mathcal L}\equiv -\bar{{\mathcal R}}_{t}=e^{\sigma/2}\left\{{1 \over 3}\cosh \left({\gamma-\delta \over 2}\right) \left[{\dot{r}^{2} \over \cosh^2\left({\gamma-\delta \over 2}\right)}-\dot{\sigma}^{2}-\dot{\delta}^{2}\right]+{2 \over 3}\sinh\left({\gamma-\delta \over 2}\right)\dot{\sigma}\dot{\delta}-V(\gamma,\delta,r^2)\right\}\label{Lsimple}
\ee

Let us recall that the Lagrangian (\ref{Lsimple}) corresponds to the restricted configuration where both particle worldlines lie in the same timelike plane. The logarithms of the  relative eigenvalues of the two metrics in this special limit are:
\be
\mu_0=2\gamma ~~~,~~~ \mu_1={2 \over 3}(\delta-\sqrt{3}r) ~~~,~~~ \mu_2={2 \over 3}(\delta+\sqrt{3}r) ~~~,~~~ \mu_3={2 \over 3}\delta
\ee

Even though up to now we have restricted the conserved quantities that appear in the system to have particular values (zero), the dynamics of the above system is still very rich and difficult to analyse. To get an insight in the behaviour of the solutions of the system, we will study two extreme cases. In the one case the two world lines $\alpha^{\mu}$ and $\beta^{\mu}$ are mostly {\textit{spacelike}} separated, and in the second   mostly {\textit{timelike}} separated. An extreme version of the first case is realized by setting $\delta=0$ and having $r\gg 1$. Similarly, an extreme version of the second case is realized in the opposite case where $r=0$ and $\delta\gg 1$. Moreover, we shall further restrict the scope of our analysis by requiring that the velocity  vectors be, at least initially, both timelike (and future directed).   We will first study the case of the Pauli-Fierz potential and then indicate the changes when considering the brane motivated one.

\section{The Pauli-Fierz potential}

The Pauli-Fierz potential (\ref{PF}) with the previous notation has the simple form:
\be
V(\gamma,\delta,r)={m^2 \over 9}\left(r^2-\delta^2-3\delta\gamma\right)\label{PFexplicit}
\ee

Let us mention the values of the energy density, the pressures as well as the equations of state for the different spatial directions in this case for the first metric:
\ba
&\rho_1=&{m^2 \over 18}e^{\gamma+\delta \over 2}(r^2-\delta^2-3\gamma \delta +6\delta)\label{den1}\\
&P_1^a=&-{m^2 \over 18}e^{\gamma+\delta \over 2}\left(r^2-\delta^2-3\gamma \delta +6\gamma+4\delta-2\sqrt{3} \{-1,1,0\} r \right)\\
&w_1^a=&{P_1^a \over \rho_1}=-1+2{\delta -3\gamma+\sqrt{3} \{-1,1,0\} r  \over r^2-\delta^2-3\gamma \delta +6\delta}\label{state1}
\ea

The corresponding quantities for the second metric are respectively:
\ba
&\rho_2=&{m^2 \over 18}e^{-{\gamma+\delta \over 2}}(r^2-\delta^2-3\gamma \delta -6\delta)\label{den2}\\
&P_2^a=&-{m^2 \over 18}e^{-{\gamma+\delta \over 2}}\left(r^2-\delta^2-3\gamma \delta -6\gamma-4\delta+2\sqrt{3} \{-1,1,0\} r \right)\\
&w_2^a=&{P_2^a \over \rho_2}=-1-2{\delta -3\gamma+\sqrt{3} \{-1,1,0\} r  \over r^2-\delta^2-3\gamma \delta -6\delta}\label{state2}
\ea

Let us now specialize in the two aforementioned limits.

\subsection{{\textit{Spacelike}} separated worldlines}
 
\begin{figure}[t]

\begin{center}
\epsfig{file=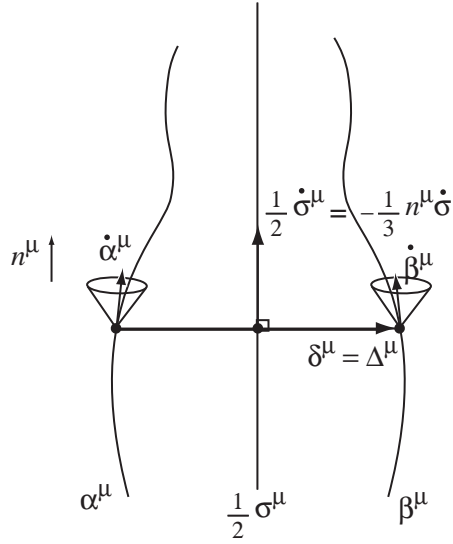}

\caption{Motion of the two ``particle'' worldlines $\alpha^{\mu}$ and $\beta^{\mu}$  for the exact  \textit{spacelike} worldline separation limit. The middle worldline is the ``center of mass'' of the system $\sigma^{\mu}/2$ and is directed along the special vector $n^{\mu}$. The motion of the two ``particles'' is symmetric with respect to the motion of their ``center of mass''.
\label{spaceplot}}
\end{center}
\end{figure}

We consider first the purest spacelike separation  corresponding to the limit $\delta=0$, where the two worldlines are symmetric  with respect to the motion of their ``center of mass'' (see Fig.\ref{spaceplot}). It is easily seen that $\delta=0$ is always a solution, if and only if one has also $\gamma=0$. The $\gamma$ and $\delta$ equations of motion are then satisfied independently of the evolution of the other degrees of freedom. The potential is simply:
\be
V(r)={m^2 \over 9} r^2 
\ee
and the Lagrangian (\ref{Lsimple}) simplifies to the rather friendly form:
\be
{\mathcal L}=e^{\sigma/2}\left({1 \over 3}\dot{r}^{2}- {1 \over 3}\dot{\sigma}^{2}-{m^2 \over 9} r^2 \right)
\ee
The above Lagrangian nicely illustrates the view of the dynamics of the system as that of two particles connected by a spring (with an energy proportional to $r^2$, {\textit{i.e.}} to the square of the spatial distance separating the two particles).

The system of equations of motion for the fields $\sigma$ and $r$  is:
\ba
{2 \over 3}\left(\ddot{\sigma}+{\dot{\sigma}^2 \over 2}\right)=V(r) \label{eq_s}\\
-{2 \over 3}\left(\ddot{r}+{\dot{\sigma}\dot{r} \over 2}\right)={\de V(r) \over \de r}\label{eq_r}
\ea
with the Hamiltonian constraint:
\be
{1 \over 3}(\dot{r}^2-\dot{\sigma}^2)+V(r)=0
\label{eq_H}
\ee
It is easy to check that by differentiating the Hamiltonian constraint and using one of the equations of motion we can obtain the remaining equation of motion. The structure of the above system is similar to the one of a scalar field in an isotropic spatially flat universe. The scalar field in the above case is $r$ while $\sigma$ is the logarithm of the scale factor. Thus, if the slow roll conditions $|\de_r V / V|^2 \ll 1$, $|\de_r^2 V / V| \ll 1$ hold  in the above system, there will be at least one {\textit{inflationary domain}}. In the particular potential at hand these two conditions can be satisfied when $r \gg 1$.

In this slow-roll region we can  find solutions of the (\ref{eq_s}), (\ref{eq_r}), (\ref{eq_H}) by omitting the second order derivatives in the equations of motion and the $\dot{r}^2$ term in the Hamiltonian constraint. The equations can then be easily integrated to:
\ba
&r=r_0-{2m \over \sqrt{3}}t\label{rsol}\\
&\sigma=\sigma_0+{m \over \sqrt{3}}\left(r_0t-{m \over \sqrt{3}}t^2\right)\label{ssol}
\ea
where the relation  $\dot{\sigma}={m \over \sqrt{3}}r$ holds. The logarithms of the  scale factors in this case for the three spatial directions are (after a rescaling of the spatial coordinates to absorb an arbitrary integration constant):
\be
\renewcommand{\arraystretch}{1.5}
\alpha^{\mu}={1 \over 2}\left\{\begin{array}{ccc}{\sigma \over 3}+{r \over \sqrt{3}}\\{\sigma \over 3}-{r \over \sqrt{3}}\\{\sigma \over 3}\end{array}\right\}~~~{\rm and}~~~\beta^{\mu}={1 \over 2}\left\{\begin{array}{ccc}{\sigma \over 3}-{r \over \sqrt{3}}\\{\sigma \over 3}+{r \over \sqrt{3}}\\{\sigma \over 3}\end{array}\right\}
\ee

Thus, from the above formulas we see that the two metrics will be exponentially inflating until the time when $r \sim {\mathcal O}(1)$. Note that the solution is written in proper time for both metrics since $\gamma=0$. The anisotropic  Hubble  parameters for the two metrics are defined as $H^{\mu}_1=\dot{\alpha}^{\mu}$ and $H^{\mu}_2=\dot{\beta}^{\mu}$ and are explicitly, along the three spatial directions:
\be
\renewcommand{\arraystretch}{1.5}
H^{\mu}_1=\dot{\alpha}^{\mu}={m \over 3}\left\{\begin{array}{ccc}{r \over 2\sqrt{3}}-1\\{r \over 2\sqrt{3}}+1\\{r \over 2\sqrt{3}}\end{array}\right\}~~~{\rm and}~~~H^{\mu}_2=\dot{\beta}^{\mu}={m \over 3}\left\{\begin{array}{ccc}{r \over 2\sqrt{3}}+1\\{r \over 2\sqrt{3}}-1\\{r \over 2\sqrt{3}}\end{array}\right\}\label{huban}
\ee
Then, the mean Hubble parameter will be the same for the two metrics and equal to:
\be
<H_1>=<H_2>={1 \over 3}\sum_{\mu}H^{\mu}_1={1 \over 3}\sum_{\mu}H^{\mu}_2={\dot{\sigma} \over 6}
\ee
Thus,  we can quantify the  anisotropy of each metric by the introduction of the parameters:
\be
A_1 \equiv \sqrt{\sum_{\mu}{(<H_1>-H_1^{\mu})^2 \over 3 <H_1>^2}}~~~;~~~A_2 \equiv \sqrt{\sum_{\mu}{(<H_2>-H_2^{\mu})^2 \over 3 <H_2>^2}}
\ee
which in the particular case at hand are:
\be
A_1=A_2=\sqrt{2} \left|{\dot{r}\over \dot{\sigma}}\right|={2\sqrt{2} \over r}\label{anspacelike}
\ee

Hence, the anisotropy in each metric {\textit{increases}} during inflation and becomes ${\mathcal O}(1)$ when the slow-roll condition is violated. This is an interesting prediction of this type of bigravity ``inflation'' which drastically differs from the scalar-inflaton-driven inflation which tends to wash out any initial anisotropies.

The equations of state for the first metric and for the three different spatial directions during the inflationary era, and their limit for $r \gg 1$ can be read from (\ref{state1}):
\be
w_1^a={P_1^a \over \rho_1}=-1 + 2\sqrt{3}(-1,1,0){1 \over r} \to -1
\ee
and for the second metric $w_2^a$ are the same with a flip in the sign in the second addendum. Note the simple relation that the equations of state for the two metrics satisfy:
\be
w_1^a+w_2^a=-2
\ee

\begin{figure}[t]

\begin{center}
\epsfig{file=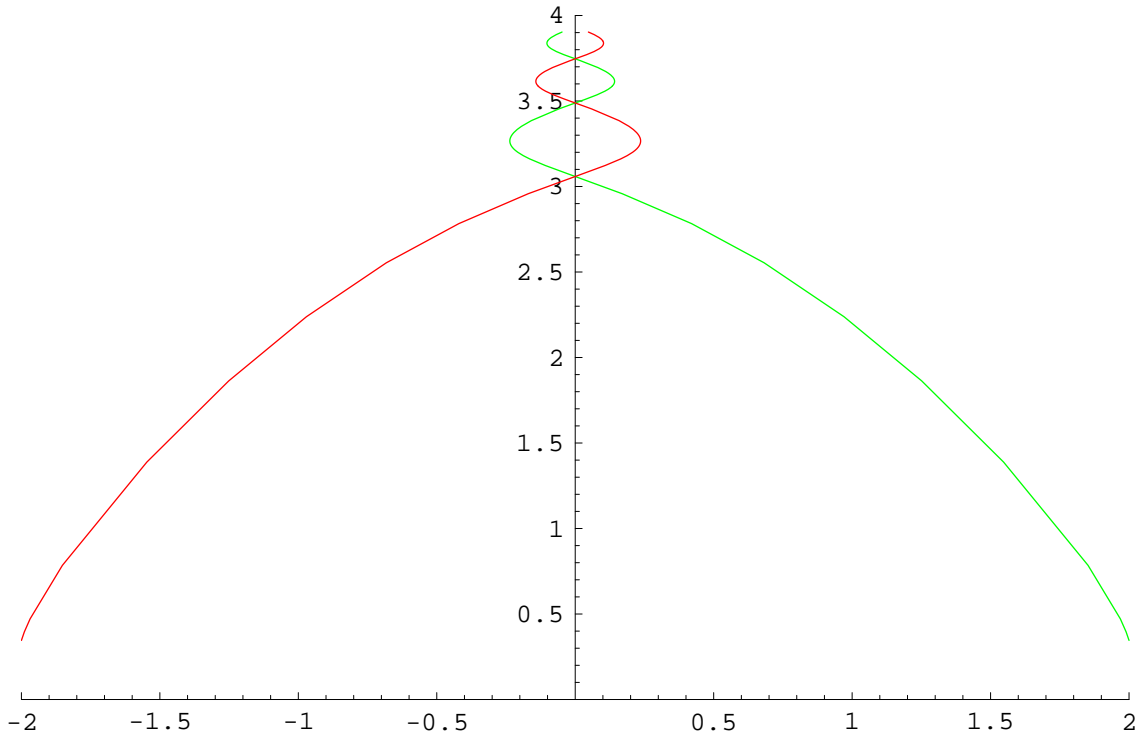}

\caption{Numerical simulation of the motion of the two ``particle'' worldlines $\alpha^{\mu}$ and $\beta^{\mu}$ for the exact  spacelike worldline separation limit and for the Pauli-Fierz potential. The horizontal axis is the distance $r/2$ from the ``center of mass'', while the vertical one is $\alpha$ and $\beta$ respectively for the two ``particles''. Initially the worldvolume of both metrics is inflating until the worldline separation becomes ${\mathcal{O}}(1)$. Then the two ``particles'' start oscillating with respect of their ``center of mass''. The oscillatory regime cannot be reached for the Pauli-Fierz potential if the timelike separation component $\delta$ is excited. \label{spaceplotnum}}
\end{center}
\end{figure}

When $r \sim {\mathcal O}(1)$ our approximation breaks down and $r$ will start to oscillate around zero. In this limit $\sigma$ grows logarithmically. The anisotropy parameter in both metrics is maximal (and ${\mathcal{O}}(1)$) near the zeros of $r$, and zero at the extrema of $r$. The evolution of the system during the inflationary and the oscillatory region is illustrated in Fig.\ref{spaceplotnum}.  However, as we are going to see, the later oscillatory region will cease to exist in the presence of  a slight initial  perturbation of $\delta$ and the relative lapse $\gamma$ will undergo (for the chosen potential) a run away towards infinity before the first zero of $r$ is reached.

Let us indeed consider the effect of perturbations  of $\delta$ away from $\delta=0$. If we let $\gamma=\delta+2\epsilon$, then the perturbation of the  Routhian in  quadratic order is:
\be
\delta {\mathcal L}=e^{\sigma/2}\left[{1 \over 3}\dot{\delta}^{2}+{\epsilon^2 \over 6}(\dot{\sigma}^{2}+\dot{r}^{2})-{2 \over 3}\epsilon \dot{\sigma}\dot{\delta}-{2 \over 9}m^2(2\delta^2+3\delta\epsilon)\right]
\ee
The constraint ${\de \delta{\mathcal L} \over \de \epsilon}=0$ can be used to solve for $\epsilon$. If we substitute it back to the Routhian we find:
\be
\delta{\mathcal L}=\underbrace{{1 \over 3}e^{\sigma/2}\left(1-{2\dot{\sigma}^{2} \over \dot{\sigma}^{2}+\dot{r}^{2}}\right)}_A\dot{\delta}^{2}-\underbrace{{m^2 \over 3}e^{\sigma/2}\left[{4 \over 3}+{2 m^2-\dot{\sigma}^2 \over \dot{\sigma}^{2}+\dot{r}^{2}}-2\left({\dot{\sigma}\over \dot{\sigma}^{2}+\dot{r}^{2} }\right)^.\right]}_B\delta^2
\label{ABdef}
\ee
The quantities $A$ and $B$ during the inflationary period and their limits for $r\gg 1$ are:
\ba
A=-{1 \over 3}e^{\sigma/2}{r^2-4 \over r^2+4}&\rightarrow& -{1 \over 3}e^{\sigma/2}\\
B={m^2 \over 3}e^{\sigma/2}\left[{4 \over 3} -{r^2-10 \over r^2+4}-{8r^2 \over (r^2+4)^2}\right]&\rightarrow& {m^2 \over 9}e^{\sigma/2}
\ea
Thus, the perturbation of the action simplifies (during slow-roll) to:
\be
\delta{\mathcal S}=\int dte^{\sigma/2}\left(-{1 \over 3}\dot{\delta}^2-{m^2 \over 9}\delta^2\right)
\ee
From the  extremization of this action we get the following motion for $\delta$:
\be
\delta=e^{-\sigma/2}\left[C_1+C_2 {\rm Erf}\left({r \over 2 \sqrt{2}}\right)\right]
\ee
where ${\rm Erf}(x)={2 \over \sqrt{\pi}}\int_0^x e^{-y^2}dy$ is the error function. Thus we see that although the field $\delta$ has an unstable looking action (after taking into account the ghost-like kinetic term of $\delta$ its correct-looking mass term corresponds in fact to a tachyonic instability), the time dependence of $\sigma$ tames and actually damps the evolution of $\delta$. This is confirmed also by numerically integrating the full equations of motion for a small $\delta$ perturbation.

However, in the region where $r \sim {\mathcal O}(1)$, where the result of the above perturbation analysis breaks down, we see numerically that $\gamma$  starts growing fast and finally, always before $r$ reaches its first zero,  it runs away to $+\infty$. In this limit we see that $\dot{\alpha}^{\mu}\dot{\alpha}_{\mu} \to 0^-$ while  $\dot{\beta}^{\mu}\dot{\beta}_{\mu} \to +\infty$. Note that $\alpha^{\mu}$ remains always timelike ($\dot{\alpha}^{\mu}\dot{\alpha}_{\mu} < 0$), but $\beta^{\mu}$ starts timelike ($\dot{\beta}^{\mu}\dot{\beta}_{\mu} < 0$), goes through its ``light cone'' ($\dot{\beta}^{\mu}\dot{\beta}_{\mu}=0$), and ends up spacelike  ($\dot{\beta}^{\mu}\dot{\beta}_{\mu} > 0$).  To understand what happens in the limit $\gamma \to +\infty$ we have examined numerically three invariant quantities for the two metrics. The first is the proper energy densities:
\be
\rho_1=-e^{\gamma}\dot{\alpha}^{\mu}\dot{\alpha}_{\mu} ~~~, ~~~ \rho_2=-e^{-\gamma}\dot{\beta}^{\mu}\dot{\beta}_{\mu}
\ee
which are proportional to the $G_0^0$ component of the Einstein tensor. The limiting behaviour of $\dot{\alpha}^{\mu}\dot{\alpha}_{\mu}$ and $\dot{\beta}^{\mu}\dot{\beta}_{\mu}$ conspire in such a way with the runaway of $\gamma$,  that the above energy densities tend to finite values, positive for $\rho_1$ and negative for $\rho_2$.  The second invariant quantity is the anisotropies of the two metrics. For the first one we observe that after having reached a maximum value, it starts decreasing, and for the second one we observe that close to the point where $\gamma$ diverges, it has a decreasing behaviour as well.  The third set of invariant quantities that one can study is the equations of state. For the first metric all three equations of state have the limit $w_1^a \to -\infty$ and for the second metric they tend to negative but finite values. From all the above we conclude that the final state of the system as measured by each metric is not singular even though the relative lapse $\gamma$ runs away to $+\infty$. 

This running away, however, signals that our effective theory description breaks down and that the limit we are discussing is governed by different dynamics. Indeed, the mass scale corresponding to $m^2$ becomes $m_1^2 \sim m^2 e^{\gamma/2}$ when viewed in the first metric and $m_2^2 \sim m^2 e^{-\gamma/2}$ when viewed from the second one  (see {\textit{e.g.}} equations (\ref{den1}), (\ref{den2})). Therefore, when $\gamma$ gets too large, this initially ``light'' scale becomes ``heavy'' on the first ``brane''\footnote{We do not have in mind any  brane configuration that can lead to  a pure Pauli-Fierz potential in its effective four dimensional action. However, we can still think of ``branes'' being  weakly coupled worlds (see \cite{DK}) in a higher dimensional setup having this particular four dimensional effective description.} and ``ultralight'' on the second. This means, for instance, that the tower of heavier graviton masses above $m_2^2$ (that are treated as infinitely heavier and have been truncated away in deriving our effective four-dimensional action) might become light and should be taken into account. The run away of $\gamma$ then signals the necessity to shift to a new basic Lagrangian.

A simple  reason of the run away of $\gamma$ in the above case can be seen from the structure of the original Lagrangian (\ref{actfix}):
\be
{\mathcal L}_t=e^{\alpha+\gamma/2}~\dot{\alpha}^{\mu}\dot{\alpha}_{\mu}+e^{\beta-\gamma/2}~\dot{\beta}^{\mu}\dot{\beta}_{\mu}-{m^2 \over 9}e^{\sigma/2}\left(r^2-\delta^2-3\delta\gamma\right)\label{actg}
\ee

Seen as a function of $\gamma$ the above Lagrangian has three important parts, namely an increasing exponential in front of the kinetic term of the one ``particle'', a falling exponential  in front of the kinetic term of the second ``particle'' and a linear contribution in the potential. If the kinetic terms of the two particles have the same sign ({\textit{i.e.}} if they are both timelike or both spacelike), there will always be an extremum of the action for a finite $\gamma$. However, when they start taking opposite signs the extremum can easily run away to infinity and disappear (except when $\delta$ has the good sign and is large enough to confine $\gamma$ in some bounded interval).

As said in Section 2 this led us to consider modified potentials, with improved ``confining'' properties for $\gamma$. For instance, an improved version of the naive Pauli-Fierz potential (\ref{PF}) is the potential (\ref{defpot}) which includes a quartic contribution (note that one generically expects the presence of such quartic contributions in nonlinear potentials $V(\mu_A)$):
\be
V(\sigma_1,\sigma_2)={m^2 \over 24}(\sigma_2-\sigma_1^2+\lambda \sigma_2^2)\label{defpot1}
\ee

Now the potential (\ref{PFexplicit})  will include in addition to the linear in $\gamma$ term, a term which is quadratic and a term which is quartic in   $\gamma$. These terms help to ensure that there exists an extremum for finite $\gamma$ and thus that the ``particle'' velocities can go through the light cone without any pathologies. Note, however, that the existence of a solution in $\gamma$ is not guaranteed for arbitrary evolutions. Indeed, the potential $V(\gamma)$, which tends, by itself, to confine $\gamma$ near $\gamma=0$, is compounded by the effect of the kinetic terms in equation (\ref{actg}) which become actually dominant for large values of $\gamma$. When $\dot{\alpha}^{\mu}\dot{\alpha}_{\mu}$ and $\dot{\beta}^{\mu}\dot{\beta}_{\mu}$ have opposite signs these kinetic-related exponential potentials might destabilize the ``local'' confining ability of $V(\gamma)$. The only case where one would be guaranteed to always have a solution in $\gamma$ is the case where a confining $V(\gamma)$ would dominate over $e^{|\gamma|/2}$ as $|\gamma| \to \infty$. Anyway, we find that the slight deformation of the potential brought by the quartic addition $\lambda \sigma_2^2$, is sufficient (for the solutions we explored) to prevent the runaway of $\gamma$ and the breakdown of the effective theory  in the pure Pauli-Fierz potential. Consequently, the oscillatory region will be present for  (\ref{defpot1}) even if $\delta$ is slightly perturbed. We have confirmed the above behaviour numerically and further observed that the maxima of the anisotropy parameter during successive oscillations are of decreasing amplitude.

\subsection{{\textit{Timelike}} separated worldlines}

\begin{figure}[t]

\begin{center}
\epsfig{file=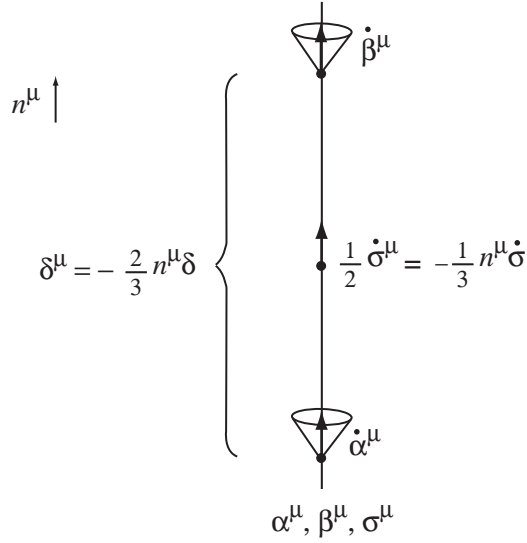}

\caption{Motion of the two ``particle'' worldlines $\alpha^{\mu}$ and $\beta^{\mu}$  for the exact  \textit{timelike} worldline separation limit. The two worldlines are collinear and the evolution of the two metrics isotropic.
\label{timeplot}}
\end{center}
\end{figure}

An extreme example of timelike worldline separation  is the limit $r=0$, where the two worldlines are collinear (see Fig.\ref{timeplot}). One checks that if initially $r=\dot{r}=0$, it remains zero all over the evolution. The potential becomes:
\be
V(\gamma,\delta)=-{m^2\over 9}\left(\delta^2+3\delta\gamma\right)
\ee
Then the Lagrangian (\ref{Lsimple}) simplifies to:
\be
{\mathcal L}=e^{\sigma/2}\left[-{1 \over 3}\cosh \left({\gamma-\delta \over 2}\right) \left(\dot{\sigma}^{2}+\dot{\delta}^{2}\right)+{2 \over 3}\sinh\left({\gamma-\delta \over 2}\right)\dot{\sigma}\dot{\delta}+{m^2\over 9}\left(\delta^2+3\delta\gamma\right)\right]
\ee

Since setting the spacelike separation between the worldlines to zero actually makes the two metrics isotropic, we can express the above Lagrangian in the following equivalent form:
\be
{\mathcal L}=-{2 \over 3}e^{\alpha+\gamma/2}\dot{\alpha}^2-{2 \over 3}e^{\beta-\gamma/2}\dot{\beta}^2+e^{\sigma/2}{m^2\over 9}\left(\delta^2+3\delta\gamma\right)
\ee

From the above it is worth noting that the $\delta^2$ term of the potential has tachyonic sign while both $\sigma$ and $\delta$ (or $\alpha$ and $\beta$) have ghostlike kinetic terms. This means that the potential $V$ has again the ``good'' relative sign allowing stable motion in such a timelike configuration. In other words, we have now two ``particles'' which are vertically separated, and which are still connected by a confining ``spring'' (with potential quadratic in the vertical separation). We expect to have solutions where the particles will chase each other, overtake each other, {\textit{etc.}} and go through some type of timelike oscillatory motion. This expected behaviour is, however, not realized in the case of the original Pauli-Fierz potential (\ref{PF}) because of the instability of the potential to confine the motion of $\gamma$. On the other hand, we have found that the expected behaviour is obtained when using the modified potential  (\ref{defpot}). Let us start by briefly describing what happens in the case of the original Pauli-Fierz potential (\ref{PF}).

The Friedmann equations (\ref{constr1a}), (\ref{constr2a}) in this case demand the positivity of the energy densities $\rho_1$, $\rho_2$. They read for this particular case:
\ba
{2 \over 3}e^{\gamma}\dot{\alpha}^2&=&-{m^2 \over 18}\delta (\delta+3(\gamma-2))e^{{\gamma+\delta \over 2}}\label{F1}\\
{2 \over 3}e^{-\gamma}\dot{\beta}^2&=&-{m^2 \over 18}\delta (\delta+3(\gamma+2))e^{-{\gamma+\delta \over 2}}\label{F2}
\ea
From these, we can easily see that there are two allowed regions in the $(\delta,\gamma)$ plane, the one with $\delta>0$, $\gamma<-2-{1 \over 3}\delta$ and the second with $\delta<0$, $\gamma>2-{1 \over 3}\delta$. Note that both allowed regions are away from  the usual ``perturbative'' domain $|\gamma| \ll 0$ (corresponding to $h^m_{MN} \ll 1$).

The solutions fall into two categories. The one which gives accelerated expansion in one of the metrics and the ones that give decelerated contraction on the same metric. Since the solutions in the above-mentioned regions can be converted to one another through a time reversal or exchange of $\alpha$ and $\beta$, it suffices to examine the first region only and in particular the expanding solutions. The motion of the system for expanding metrics in this case has the following behavior in the $(\delta,\gamma)$ plane:   initially $\gamma$ increases with time, while $\delta$ is decreasing. If we evolve the solution back in time, we hit a cosmological initial singularity, where $\gamma \to -\infty$, $\delta \to \infty$. If we evolve forward in time, we find that, after $\delta$ has decreased to the  region where  $\delta \sim {\mathcal{O}}(1)$, $\gamma$ reaches a maximum value and starts decreasing, while $\delta$ keeps decreasing until $\delta \to 0$. At this latter limit $\gamma \to -\infty$. In both regions $\sigma$ is increasing in such a way that $\beta$ is very slowly varying being in very good approximation constant ({\textit{i.e.}} the second metric is approximately flat). On the other hand, the first metric in the region of $\delta \gg 1$ experiences a power-law accelerated expansion, which is converted to deflation in the limit $\delta \to 0$.  Details of these solutions are given in Appendix C.

Let us now describe the behaviour of our solutions obtained when working with the ``nicer'' potentials, able to confine the motion of $\gamma$, such as the modified Pauli-Fierz potential (\ref{defpot1}). In that case we numerically found that the presence of terms in the potential $V(\gamma)$ quadratic and quartic in $\gamma$ are sufficient to bound the motion of $\gamma$ in a finite range. We then  insert the solution for $\gamma$ of ${\de {\cal L} \over \de \gamma}=0$ into the evolution equations for the other ``timelike'' degrees of freedom, {\textit{i.e.}} for $\alpha$ and $\beta$, or equivalently for $\sigma=\alpha+\beta$ and $\delta=\beta-\alpha$. A numerical study of the evolution of $\gamma$, $\sigma$ and $\delta$  then shows that the expected behaviour of the two vertically separated ``particles'' connected by a ``spring'' essentially holds true, with, however, an important effect linked to the ``center of mass'' motion. If, for instance, we consider the case where initially some ``particle'', say $\alpha^{\mu}$, is ``below'' the other ``particle'' $\beta^{\mu}$, we found (for the initial velocities $\dot{\alpha}$, $\dot{\beta}$ we considered) that the ``lagging'' ``particle'' $\alpha^{\mu}$ starts to chase the ``leading'' one, and tries to overtake it. In other words, if $\delta=\beta-\alpha=n_{\mu}(\beta^{\mu}-\alpha^{\mu})$ is initially negative\footnote{Remember from Fig.\ref{timeplot} that $\delta$ is opposite to the vertical separation between $\beta^{\mu}$ and $\alpha^{\mu}$ because of the timelike character of $n^{\mu}$.}, it will tend to increase.

However, we found that the further evolution of $\delta$ disagrees with the naive expectation of a timelike harmonic oscillator. Indeed, we found that $\delta$ initially increases, but then stabilizes, after some damped oscillations, around a negative final value without ever crossing zero. [Correspondingly $\gamma$ also stabilizes,  after some damped oscillations,  around some final value.] This behaviour is clearly due to the coupling between the evolution of $\delta$ and the evolution of $\sigma$, {\textit{i.e.}} the coupling between the ``relative motion'' of the pair of ``particles'', with its ``center of mass'' motion. The latter  tending to damp the former, and to maintain it away from zero, {\textit{i.e.}} away from a zero-separation configuration. We expect such a behaviour to be generic, though its details might depend on the chosen initial data. [For instance, for some well chosen initial velocities the lagging particle should be able to overtake the leading one so that one could have one or a few vertical oscillations before the particles stabilize into a permanently ``chasing'' configuration with non-zero vertical separation.]

What is physically interesting in this behaviour is that we have here a {\textit{locking mechanism}} by which the coupled evolution of isotropic metrics $\chi^1_{\mu \nu}=e^{2\alpha/3}\delta_{\mu \nu}$, $\chi^2_{\mu \nu}=e^{2\beta/3}\delta_{\mu \nu}$, {\textit{i.e.}} a ``bi-Friedmann'' configuration, locks itself up, after some damped oscillations, in a stable  ``chasing'' configuration where $\delta=\beta-\alpha$ is constant, as well as $\gamma$. In view of the zero energy constraint (\ref{constr1b}), taken in the limit where $\dot{\delta}^{\mu}=0$ and $\dot{\sigma}^{\mu}\dot{\sigma}_{\mu}=-{2 \over 3}\dot{\sigma}^2$,  the final configuration must be such that:
\be
{1 \over 3}\cosh \left({\gamma-\delta \over 2}\right)\dot{\sigma}^2=V(\gamma,\delta)\label{rate}
\ee

Therefore, $V$ must end up being {\textit{positive}}, and $\dot{\sigma}^2$ ends up being a constant. In other words, the locking of the two metrics fixes the value of the bimetric potential to a positive value which then behaves as a usual cosmological constant term driving an exponential inflation, at the same rate (as measured in the average proper time) $\dot{\alpha}=\dot{\beta}={\dot{\sigma} \over 2}$, for the two metrics. Note, however, that the physical values of the two expansion rates, as measured in each corresponding metric, are different: $H_1=e^{\gamma/2}H_0$ and $H_1=e^{-\gamma/2}H_0$ where $H_0=\dot{\sigma}/ 6$.  In view of the potential physical importance of such a ``bi-de-Sitter'' locked configuration, let us consider the general conditions for such a stationary configuration to exist. 

Starting, for instance, from the full set of ``timelike'' equations given in Appendix A, one easily finds that the necessary and sufficient conditions for a locking solution ({\textit{i.e.}} $\gamma=$const., $\delta=\beta-\alpha=$const., $\dot{\alpha}=\dot{\beta}={\dot{\sigma} \over 2}=$const.) to exist are:
\ba
{\de V(\gamma,\delta) \over \de \gamma}={\de V(\gamma,\delta) \over \de \delta}\label{dgcond}\\
{1 \over V}~{\de V(\gamma,\delta) \over \de \gamma}=-{1 \over 2}\tanh \left({\gamma-\delta \over 2}\right)\label{dcond}
\ea
with the additional requirement that $V$ is positive so that (\ref{rate}) can hold. Then the constant common rate of expansion $\dot{\alpha}=\dot{\beta}={\dot{\sigma} \over 2}$ is given by Eq.(\ref{rate}). Note that Eqs.(\ref{dgcond}), (\ref{dcond}) represent two equations for two unknowns. One generically expect this system to admit solutions. If we consider the class of potentials that depend only on $\sigma_1$ and $\sigma_2$, $V=V(\sigma_1,\sigma_2)$, using (\ref{s1+v}), (\ref{s2+v}) with $r=0$, {\textit{i.e.}}:
\be
\sigma_1=2(\gamma+\delta)~~~,~~~\sigma_2=4\left(\gamma^2+{1 \over 3}\delta^2\right)
\ee
one easily sees that the condition (\ref{dgcond}) simplifies to:
\be
\left(\gamma-{\delta \over 3}\right){\de V \over \de \sigma_2}=0\label{lockcond}
\ee

There are two types of solutions of the above equation: either $\delta=3\gamma$ or $\de_{\sigma_2}V=0$. Note that we already encountered the condition  $\de_{\sigma_2}V=0$ in Appendix B as the necessary condition for nontrivial ``symmetry breaking'' effects to occur ({\textit{i.e.}} that $V$, considered as a function of the relative shift vector $\vec{b}$, be roughly of the symmetry breaking form $V=V_0-\vec{b}^2+\vec{b}^4$). However, the potentials that we are currently considering (including their $\lambda$-modified versions, {\textit{e.g.}} (\ref{defpot1})) do not admit ``critical points'' where  $\de_{\sigma_2}V=0$, if $\lambda > 0$. Let us then concentrate on the other type of universal solution of the constraint (\ref{lockcond}), namely:
\be
\delta=3\gamma\label{lockgd}
\ee

Inserting this general solution in (\ref{dcond}), we then get one constraint for one unknown (say $\gamma$). Note, however, that not all solutions of this constraint lead to locking solutions. Indeed, one must further require the corresponding value of $V$ to be {\textit{positive}} so that one indeed ends up with a (real) ``bi-de-Sitter'' solution with common expansion rate:
\be
\dot{\alpha}^2=\dot{\beta}^2={3 \over 4}{V \over \cosh \left({\gamma-\delta \over 2}\right)}\label{lockedrates}
\ee

The solutions satisfying (\ref{lockgd}) lead to the following metrics:
\begin{equation}
\renewcommand{\arraystretch}{1.5}
\begin{array}{c}
ds_1^2=e^{-\gamma}ds_0^2 \\ 
ds_2^2=e^{+\gamma}ds_0^2
\end{array}
 ~~~~~{\rm where}~~ ds_0^2=-dt^2+e^{\sigma/3}\delta_{\mu \nu} dx^{\mu}dx^{\nu}
\end{equation}
Note again that the physical expansion rates are $H_1=e^{\gamma/2}H_0$ and $H_1=e^{-\gamma/2}H_0$ where $H_0=\dot{\sigma}/ 6$. One can check that such bi-de-Sitter solutions in bigravity admit, for general classes of couplings, generalizations to multi-de-Sitter solutions in  multigravity theories with $N$ coupled metrics. The conditions for this to happen is that $\Gamma_i=\Delta_i/3$, where  $\Gamma_i=\gamma_{i+1}-\gamma_i$ and $\Delta_i=\alpha_{i+1}-\alpha_i$, with $i=1,\dots,N-1$ ($\gamma_i$, $\alpha_i$ being the logarithmic time shifts and scale factors respectively for the $i$-th metric) and additionally that there exist real solutions to the $N-1$ equations (similar to (\ref{dcond})) that determine the $\Gamma_i$. Then one indeed finds that in the appropriately defined averaged proper time coordinate:
\be
ds_i^2=e^{f_i}ds_0^2 ~~~~, {\rm where}~~ ds_0^2=-dt^2+e^{\sigma/3}\delta_{\mu \nu} dx^{\mu}dx^{\nu}\label{multilock}
\ee
with $\sigma=\alpha_1+\alpha_N$ and $\displaystyle{f_i=\sum_{k=1}^i \Gamma_k-\sum_{k=i+1}^{N}\Gamma_k}$ .

For instance, for the original Pauli-Fierz potential, Eq.(\ref{dcond}) with (\ref{lockgd}) leads to the constraint:
\be
\gamma \tanh \gamma =1
\ee
which has the two solutions $\gamma_c \approx \pm 1.20$. However, the corresponding value of $V$ ($V=-2m^2\gamma_c^2$) is negative. By contrast, we find that the modified Pauli-Fierz potential (\ref{defpot1}) with the quartic term $\lambda \sigma_2^2$ admits locking solutions. In the latter case,  Eq.(\ref{dcond}) with (\ref{lockgd}) leads to the constraint:
\be
\gamma \tanh \gamma = {{16 \over 3}\lambda \gamma^2-1 \over {8 \over 3}\lambda \gamma^2-1}
\ee
which has for any $\lambda > 0$ two pairs of roots (each pair consisting of two opposite roots), one pair with ${8 \over 3}\lambda \gamma_c^2-1 < 0$ (which tends to $\gamma_c \approx \pm 1.20$ as $\lambda \to 0$) and another pair with  ${8 \over 3}\lambda \gamma_c^2-1 > 0$ (which behaves as $\gamma_c \approx \pm \left({1 \over \sqrt{\lambda}}+{1 \over 2}\right)$ as $\lambda \to 0^+$). For $\lambda < 0$ there is only one pair of opposite roots. The value of the potential, on the other hand, is:
\be
V=2m^2\gamma_c^2\left({8 \over 3}\lambda \gamma_c^2-1\right)
\ee
which is positive only if $\lambda > 0$ and additionally ${8 \over 3}\lambda \gamma_c^2-1 > 0$. Thus, the requirement of positivity of the potential excludes the $\lambda < 0$ case and keeps only one pair of solutions for the  $\lambda > 0$ case. Note that {\textit{any}} positive value  $\lambda$ would give a locking solution. 

Summarizing, general classes of potentials admit timelike separated configurations which lock in a stable\footnote{We have not investigated the general conditions for stability of such locked configurations when they exist. However, they seem to be numerically stable even when introducing a tilt angle away from the vertical direction.} inflationary regime corresponding to a ``bi-de-Sitter'' configuration, with the same expansion rate. We see that bigravity can therefore play the role of dark energy in driving cosmic acceleration. But, contrary to the spacelike separated case (which led to an anisotropic type of slow-roll inflation, followed by power law expansion driven by spacelike oscillators) the pure timelike separated case is consistent with an isotropic de-Sitter-like inflation. When comparing such a final behaviour, in the modified Pauli-Fierz case, with the discussion of the original Pauli-Fierz case in Appendix C, it is striking to note how a simple (and natural) modification of the potential can drastically affect the set of bigravity solutions. This is a clear illustration of the concept of ``universality classes'' of bigravity potentials discussed in \cite{DK}.

It is interesting to compare our locked solutions to the accelerating solution found in \cite{Deffayet:2000uy,Deffayet:2001pu} within the context of brane-induced gravity. In the late time limit the solution of \cite{Deffayet:2000uy,Deffayet:2001pu} is given by:
\be
ds^2=dy^2+(1+|y|)ds_0^2 ~~~~, {\rm where}~~ ds_0^2=-dt^2+e^{2Ht}\delta_{\mu \nu} dx^{\mu}dx^{\nu}\label{def}
\ee
The fact that the four-dimensional part of the above five-dimensional metric is conformal (with a $y$-dependent conformal factor) to a fixed de-Sitter metric is similar to the particular case (\ref{multilock}) of our general locked solutions. Eq. (\ref{def}) can be easily understood as following. The five-dimensional Einstein action contains the term $\int \sqrt{-g}[tr(g^{-1}\de_y g)^2-(tr g^{-1}\de_y g)^2]$. This term can be viewed as the continuum limit of an infinite sum of ``nearest neighbour'' interactions $\sum {\mathcal V}(g_{i+1}^{-1}g_i)$ (see \cite{DK}) when considering the $N \to \infty$ limit of multigravity (where $N$ is the number of coupled metrics). Based on this reinterpretation of the particular five-dimensional model studied in  \cite{Deffayet:2000uy,Deffayet:2001pu}, we can view it as the $N \to \infty$ limiting case of the general class of de-Sitter locked solutions mentioned above.

\section{The brane motivated potential}

The brane motivated potential (\ref{branepot}) with the previous notation reads:
\be
V(\gamma,\delta,r)=m^2\left[\cosh \left({1 \over 2 \sqrt{3}} \sqrt{{8 \over 3}r^2+{1 \over 3}\delta^2+3\gamma^2-2\delta\gamma}\right) - \cosh \left({\delta+\gamma \over 2}\right)\right]\label{branepotexpl}
\ee

\subsection{Absence of timelike separated worldline solutions}

If we look for solutions where the two worldlines are purely timelike separated ($r=0$), the Lagrangian (\ref{Lsimple}) becomes:
\be
{\mathcal L}=e^{\sigma/2}\left[-{1 \over 3}\cosh \left({\gamma-\delta \over 2}\right) \left(\dot{\sigma}^{2}+\dot{\delta}^{2}\right)+{2 \over 3}\sinh\left({\gamma-\delta \over 2}\right)\dot{\sigma}\dot{\delta}-V(\gamma,\delta)\right]
\ee
with the potential:
\be
V(\gamma,\delta)=m^2\left[\cosh \left({\gamma-\delta/3 \over 2} \right) - \cosh \left({\gamma+\delta \over 2}\right)\right]
\ee

In this case, one finds that the two constraints (\ref{constr1a}), (\ref{constr2a}) force $\delta=\gamma=0$. Thus, the only solution is $\dot{\sigma}=0$. In other words,  there are no nontrivial solutions with $r=0$. Note that this does not preclude that the existence of timelike separated solutions with non-vertical separation, {\textit{i.e.}} with $\delta^{\mu}$ timelike but not parallel to $n^{\mu}$. Note also that there is no contradiction between the facts that: (i) the brane potential reduces to the Pauli-Fierz one for small $\gamma$ and $\delta^{\mu}$, (ii) the Pauli-Fierz potential admits purely vertical timelike evolutions, and (iii) the brane potential admits no such evolutions. Indeed, the Pauli-Fierz timelike evolutions only exist when $\gamma$ is of order unity.

\subsection{Spacelike separated worldline  solutions}

In the opposite regime where the two worldlines are required to be purely spacelike separated ($\delta= 0$) one again finds that the only way to freeze $\delta$ to zero is to have also $\gamma=0$. Then the  Lagrangian (\ref{Lsimple}) becomes:
\be
{\mathcal L}=e^{\sigma/2}\left({1 \over 3}\dot{r}^{2}- {1 \over 3}\dot{\sigma}^{2}-V(r) \right)
\ee
with the potential:
\be
V(r)=m^2\left[\cosh \left({\sqrt{2} \over 3}r \right) - 1\right]
\ee

Let us note here that the exponentially growing structure of the above potential (with a coefficient of $r$ of order unity),  makes us expect that there will be no slow-roll region. We can (at best) expect that such an exponential ``spring'' will lead to power-law inflation.   Asymptotically, for $r \gg 1$ the action can be written as:
\be
{\mathcal S}=\int dt e^{\sigma/2}\left({1 \over 3}\dot{r}^{2}-{1 \over 3}\dot{\sigma}^{2}-{m^2 \over 2}e^{{\sqrt{2} \over 3}r}\right)=\int d\tau \left({1 \over 3}r^{\prime 2}-{1 \over 3}\sigma^{\prime 2}-{m^2 \over 2}e^{\sigma+{\sqrt{2} \over 3}r}\right)
\ee
where we have changed our time variable to $d\tau={dt \over e^{\sigma/2}}$. Defining new variables as:
\be
\xi_1=\sigma+{\sqrt{2} \over 3}r~~~,~~~\xi_2={\sqrt{2} \over 3}\sigma+r \label{x1x2rs}
\ee
we can have a considerable simplification in our action:
\be
{\mathcal S}=\int d\tau \left[{3 \over 7}\left(\xi_2^{\prime 2}-\xi_1^{\prime 2}\right)-{m^2 \over 2}e^{\xi_1}\right]
\ee

This ``Toda model'' is integrable because the $\xi_2$ variable is ignorable. We have:
\be
\xi_2=\xi_{20}+p\tau  \label{x2sol}
\ee
where $p$ is the conserved momentum of $\xi_2$ and $\xi_{20}$ an integration constant. From the Hamiltonian constraint we immediately obtain for $p \neq 0$:
\be
\xi_1=-\log\left[{7m^2 \over 6p^2} \sinh^2\left({p\tau \over 2}\right)\right] \label{x1sol}
\ee
where we have set an integration constant, which accounts for time shifts, to zero.  For the special case of $p=0$ we get:
\be
\xi_1=-\log\left[{7m^2\tau^2 \over 24}\right]
\ee

There is an obvious symmetry in the simultaneous reflections $\tau \to -\tau$ and $p \to -p$. These map solutions of expanding volume (increasing $\sigma$) to ones of shrinking one. We will be interested in expanding volume solutions where $\tau <0$.  The parameter $p$ represents a kind of initial kinetic energy. Let us focus here on the simplest case $p=0$, which corresponds to the usual rolling-type inflationary solution with kinetic energy small compared to the potential energy. See Appendix D for a discussion of the general case $p \neq 0$.

For this case ($p=0$) we have for any value of the  time parameter $\tau$:
\be
\sigma = -{18 \over 7}\log |m\tau| + {\rm const.} ~~~,~~~r = -{\sqrt{2} \over 3}\sigma + {\rm const.}\label{p=0tau}
\ee
The proper time $t$ (which is proportional to the individual proper times $t_1,t_2$) is related to $\tau$ via $t \propto \tau^{-2/7}$, so that the solution (\ref{p=0tau}) reads in proper time:
\be
\sigma = 9\log (mt) + {\rm const.} ~~~,~~~r = -3\sqrt{2}\log (mt) + {\rm const.}\label{p=0t}
\ee

By rescaling coordinates to absorb the above constants we finally see that, in conformity with naive expectations, each metric has a {\textit{power-law behaviour}}, and if we parametrise it in the standard way as:
\be
ds^2=-dt^2+\sum_{\mu=1}^3(m t)^{2 p^{\mu}}(dx^{\mu})^2
\label{Kas}
\ee
we have that the Kasner exponents $p^{\mu}$  are:
\be
\renewcommand{\arraystretch}{1.5}
p^{\mu}_{(1)}=\left[\begin{array}{ccc}{3 \over 2}-\sqrt{{3 \over 2}}\\{3 \over 2}+\sqrt{{3 \over 2}}\\{3 \over 2}\end{array}\right]~~~{\rm and}~~~p^{\mu}_{(2)}=\left[\begin{array}{ccc}{3 \over 2}+\sqrt{{3 \over 2}}\\{3 \over 2}-\sqrt{{3 \over 2}}\\{3 \over 2}\end{array}\right]
\ee
We should stress here that, when we bring one of the metrics to the above form (\ref{Kas}) by absorbing the various constants in $x^{\mu}$ redefinitions, the second one will not have the form (\ref{Kas}). Thus the exponents we have written above are in the coordinate system where each of the metric (but not both) simplifies. The above exponents do not satisfy the quadratic (zero-mass-shell) Kasner relation, since  $\sum_{\mu} (p^{\mu})^2-\left(\sum_{\mu} p^{\mu}\right)^2=-{21 \over 2}$. In addition, $\sum_{\mu} p^{\mu}={9 \over 2}$, which means that each metric's volume $v_i$ expands as a function of its respective proper time as $v_i \propto t_i^{9/2}$, with $i=1,2$. Note that the latter volume expansion has an accelerating behaviour, but as expected, much less pronounced than the case of the polynomial Pauli-Fierz potential which led to exponentially inflating solutions for such a rolling initial state.   The evolution is  highly anisotropic, with the anisotropy parameter being:
\be
A_1=A_2={2 \over 3}
\ee

In all the above discussion it is always assumed that $r \gg 1$ when the simplified action is valid. For  $r \ll 1$ the potential is identical with the Pauli-Fierz one and the $r$, $\sigma$ will behave exactly as we have discussed in the previous section for this case.

Let us now check again if the obtained solution is stable against perturbations of $\delta$. We have to recalculate the quantities $A$ and $B$ of Eq.(\ref{ABdef})  and their limits for  $r \gg 1$. They read:
\be
A=-{7 \over 33}e^{\sigma/2} ~~~,~~~B\to {2m^4 t^2 \over 297}e^{\sigma/2}
\ee
Thus, the variation of the action simplifies to:
\be
\delta{\mathcal S}=\int dte^{\sigma/2}\left(-{7 \over 33}\dot{\delta}^2-{2m^4t^2 \over 297}\delta^2\right)
\ee
From the extremization of this action, we get the following motion for $\delta$:
\be
\delta={1 \over t^{7/4}}\left[C_1 I_{7/8}\left({1 \over 2}\sqrt{2 \over 63}m^2t^2\right)+C_2 K_{7/8} \left({1 \over 2}\sqrt{2 \over 63}m^2t^2\right)\right]
\ee
where $I_n(x)$ and $K_n(x)$ are the modified Bessel functions of first and second kind respectively. Hence the solution is unstable at late times. This instability is again linked to a run away of $\gamma$ towards large values (for which the above perturbative treatment breaks down). Numerically we observe that as soon as $|\gamma|$ starts to increase, it tends to run away very quickly towards infinity. The behaviour in this limit is similar to the one presented for the Pauli-Fierz potential.
 
Having understood that the brane potential (\ref{branepotexpl}) is, like the Pauli-Fierz potential, not efficient enough\footnote{This is a priori a bit surprising because $V_{\rm brane}(\gamma)$ increases exponentially as $|\gamma| \to \infty$. However,  $V_{\rm brane}(\gamma)$ increases no faster than the kinetic terms in Eq.(\ref{actg}), {\textit{i.e.}} like $a e^{\gamma/2}+b e^{-\gamma/2}$, and the asymptotic coefficients $a$ and $b$ depend on the value of $\delta$. Therefore, it it can become subdominant with respect to the (sign-changing) kinetic term contributions.} in confining $\gamma$ in the long term, we have also numerically studied the cosmological evolutions obtained with a brane potential modified by the addition of a term $\propto \lambda \sigma_2^2$. Our (partial) numerical study then shows that such a better-confining brane-type potential leads to long-surviving solutions which are qualitatively similar to the ones discussed above (in the modified Pauli-Fierz case). Again, we find that the simple picture of two ``particles'' connected by a ``spring'' is a good qualitative guide. We found spacelike separated solutions which start, as above with a power-law anisotropic expansion, and which end up in some (power-law expanding) ``spatially'' oscillating regime. The modification also allows now for purely timelike separated solutions which end up, after ``locking'', in a bi-de-Sitter final configuration.

\section{Transition between matter domination and vacuum bigravity domination}

In the previous sections we have seen that, for both potentials that we have studied, there exists a period of acceleration for one or both of the metrics. It is tempting to suggest that this acceleration  due to the metric coupling potential can be identified with the current acceleration of the universe. Let us see if this type of ``dark energy'' can be phenomenologically acceptable. A basic ingredient of any theoretical model of dark energy is that there exist a smooth transition between a matter dominated universe and a dark energy dominated one. To discuss this transition for our bigravity system we consider the addition of matter densities in the action (\ref{actnofix}), having a scaling law $\rho_{1m}=\rho_{1m0}e^{-\alpha}$ for the first metric and $\rho_{2m}=\rho_{2m0}e^{-\beta}$ for the second one, where $\rho_{1m0}$ and $\rho_{2m0}$ are some initial values of the two energy densities:
\be
{\mathcal L}_0=e^{\alpha-\gamma_1}~\dot{\alpha}^{\mu}\dot{\alpha}_{\nu}+e^{\beta-\gamma_2}~\dot{\beta}^{\mu}\dot{\beta}_{\nu}-e^{(\alpha+\beta+\gamma_1+\gamma_2)/2}V(\gamma,\delta^{\mu})-e^{\gamma_1}\rho_{1m0}-e^{\gamma_2}\rho_{2m0}
\label{actnofixmatter}
\ee
The two Friedmann equations (in the $\bar{\gamma}=0$ gauge) are modified as: 
\ba
e^{\gamma}\dot{\alpha}^{\mu}\dot{\alpha}_{\mu}+\rho_1+\rho_{1m}=0 \\
e^{-\gamma}\dot{\beta}^{\mu}\dot{\beta}_{\mu}+\rho_2+\rho_{2m}=0
\ea
where, keeping the same notation as in the previous sections, we denote by $\rho_1$ and $\rho_2$ the massive graviton dark energy densities. 

We have done some numerical experiments on this transition, notably in the case of the quartically-modified Pauli-Fierz potential (\ref{defpot}). In the examples we studied the confining power of $V(\gamma)$ was strong enough to keep $\gamma$ in some finite range, even when the potential contribution in Eq.(\ref{actnofixmatter}) is very small compared to the other terms. Then we have observed numerically that there was a smooth transition between matter domination and dark energy domination for both the spacelike and the timelike worldline separation cases.

Rather than describing in detail these numerical results (which are not particularly illuminating) let us explain with the help of simple analytical formulas what, we think, is the general simple behaviour of this transition in bigravity theories. Note first that several types of transitions can occur. If we are interested in the phenomenology on the ``first brane'', $g^1_{MN}$, either $\rho_1$ overtakes $\rho_{1m}$ before $\rho_2$ overtakes $\rho_{2m}$, or the reverse. Our (partial) numerical experiments indicate that these different  transitions lead to the same phenomenological features on the first brane. For simplicity, let us then consider the case where $\rho_1$ overtakes $\rho_{1m}$ before $\rho_2$  overtakes $\rho_{2m}$. Then, during matter domination (on both branes), both $\chi^1_{\mu \nu}$ and $\chi^2_{\mu \nu}$ expand as $t^{2/3}$ (as $\gamma$ is fixed the average proper time $t$ is proportional to both ``intrinsic'' proper times). This means that $\chi^1_{\mu \nu}(t)=a_{\mu \nu} t^{2/3}$,  $\chi^2_{\mu \nu}(t)=b_{\mu \nu} t^{2/3}$. Here, $a_{\mu \nu}$ and  $b_{\mu \nu}$ are some constant tensors. Each one can be reduced (by a change of spatial coordinates) to the normal (flat) isotropic Friedmann form, say $a'_{\mu \nu}=\delta_{\mu \nu}$. However, it is important to realize that the spatial transformation $x^{\prime \mu}=\Lambda^{\mu}_{\nu}x^{\prime \nu}$ which will ``isotropize''   $\chi^1_{\mu \nu}$, will, in general, leave  $\chi^2_{\mu \nu}$ in a general, apparently anisotropic form $b'_{\mu \nu} \neq \lambda \delta_{\mu \nu}$. In fact the ``difference'' between  $a_{\mu \nu}$ and  $b_{\mu \nu}$, {\textit{i.e.}} more precisely the matrix $(\chib^{-1}_1\chib_2)^{\mu}_{\nu}=a^{\mu \kappa}b_{\kappa \nu}$ can a priori take arbitrary values. The corresponding logarithmic eigenvalues $\mu_A=\log ({\rm eigenvalues}({\bf a}^{-1}{\bf b}))$ essentially define an ``initial'' field-space vector $\delta^{\mu}=\beta^{\mu}-\alpha^{\mu}$ separating the two ``particles'' during matter domination. In other words, in the mechanical language of the model of Fig.\ref{genplot}, during matter domination (where the ``spring'' between the two worldlines has only a subdominant effect\footnote{Note, however, that, even when $V$ is negligible with respect to the other terms, its presence remains crucial for pinning $\gamma$ to some specific value, which we assume here to remain ${\cal O}(1)$.}) the separation vector between the two ``particles'' remains constant: $\delta^{\mu}=\beta^{\mu}-\alpha^{\mu}=$const. Under our assumption about the ability of $V(\gamma)$ to confine $\gamma$, we therefore see that, during matter domination (or for that purpose, also radiation domination, where $\chi^1_{\mu \nu}(t)=a_{\mu \nu} t^{1/2}$, {\textit{etc.}}) the effective vacuum energy seen by each metric, {\textit{e.g.}}   $\rho_1=e^{(\gamma+\delta)/2}({1 \over 2}V(\gamma,\delta^{\mu})-\de_{\gamma}V(\gamma,\delta^{\mu}))$, remains essentially {\textit{constant}}. It plays therefore the same role (modulo possible $\gamma$-modulations) as a cosmological constant, during any period where $V$ is subdominant.

However, when, because of the decrease of $\rho_{1m} \propto e^{-\alpha}$ during expansion, $\rho_1$ starts dominating over $\rho_{1m}$ the situation will change and the separation vector $\delta^{\mu}$ will start to vary. In fact, if we are in the case where the ``initial'' value of $\rho_1$ (as determined by the nearly constant values of $\gamma$ and $\delta^{\mu}$ during matter domination) is {\textit{positive}} we simply expect that $\rho_1$ will take over $\rho_{1m}$ the role of driving the expansion. Such a situation can a priori occur for any ``spacetime direction'' of the connecting vector $\delta^{\mu}$. It is here that a bigravity origin of dark energy (``tensor quintessence'') can lead to interesting phenomenological predictions that differ from the standard (scalar-quintessence-like) models. 

Two contrasting cases can occur, depending on the timelike versus spacelike character of the initial connecting vector $\delta^{\mu}$. First, let us consider the case where $\delta^{\mu}$ is timelike, and even, for simplicity, of the extreme timelike type $\delta^{\mu} \propto \delta n^{\mu}$, {\textit{i.e.}} $r=0$. In this case, if there exists a stable locking configuration (as described above), we expect $\delta$ to ultimately stabilize (after some oscillations and after the second brane transits to a vacuum dominated state) in this locked state: $\delta=3\gamma=$const. Note that, in the intermediate stage where the second brane is still matter dominated there will be an  ``external force'' acting on $\delta=\beta-\alpha$ (through the equation of motion of $\beta$) so that one expects $\delta$ to deviate, in a time-dependent way, from the final locked configuration. In other words, we expect in this case  the effective equation of state of dark energy: 
\be
w_1={P_1 \over \rho_1}=-{V-2\de_{\delta} V \over V-2\de_{\gamma} V}
\ee
to depend on time, before ultimately stabilizing into the standard ``vacuum'' value $w_1=-1$ corresponding to the final, bi-de-Sitter locked configuration.

Let us now consider the other possible case where $\delta^{\mu}$ is spacelike, and even, for simplicity, of the extreme spacelike case where $\delta^{\mu} \perp n^{\mu}$, {\textit{i.e.}} $\delta=0$. In this case, an interesting prediction of bigravity is that the expansion rates of each universe (which, during matter domination were {\textit{both}} intrinsically isotropic) will start to become anisotropic. Indeed, we expect that (at least after the second universe also transits to a vacuum dominated state) the solution for the connecting vector $\delta^{\mu}$ (which started with the initial conditions $\delta^{\mu}=\delta_0^{\mu}$, $\dot{\delta}^{\mu}=0$) will quickly be attracted towards the spacelike slow-roll solutions described in Section 5. This means that, within one Hubble time or so after the transition, the intrinsic expansion rates along the three special spatial directions (corresponding to the eigenvectors of the matrix ${\bf a}^{-1}{\bf b}$) will differ from each other. We expect, however, like in Eqs.(\ref{huban})-(\ref{anspacelike}) above (which are not directly applicable because they were derived for the original Pauli-Fierz potential) that, if the magnitude $r=\sqrt{{3 \over 2}\Delta^{\mu}\Delta_{\mu}}$ of the (purely spacelike) connecting vector $\delta^{\mu}=\Delta^{\mu}$ is large  compared to one, the spatial anisotropies in the rates of expansion will be rather small. As for the time dependence of the expansion, it will be initially of the slow-roll inflationary type (as in Eq.(\ref{rsol}),(\ref{ssol})), but will ultimately end up as a power-law-type expansion as $r$ becomes ${\cal O}(1)$ and starts oscillating around zero.

We have checked the occurrence of the two types (timelike versus spacelike) of transition to bigravity vacuum dominance in some numerical experiments. We have, however, not tried to make exhaustive experiments because there are many parameters that can influence the actual time-dependence of the transition: the shape of the potential, its ability to confine $\gamma$ or not, the value of the the ratio between $\rho_{1m}$ and $\rho_{2m}$, the presence or absence of a ``locked'' solution, {\textit{etc.}} It is too early to embark on such an exhaustive study. Our main purpose in this work was only to delineate some of the qualitative new features of possible bigravity origin of cosmic acceleration.

\section{Conclusions}

We have explored general spatially-flat cosmological solutions (of the anisotropic Bianchi I type) of classes of nonlinear bigravity theories. Even within this restricted class of homogeneous cosmologies we focused on special cases. We did not explore the possibility (mentioned in Section 3) where, due to a type of spontaneous symmetry-breaking mechanism, the relative shift vector $b^{\mu} \equiv e^{-\bar{\gamma}}(b_2^{\mu}-b_1^{\mu})$ be non-zero. After a brief discussion of the structure of the evolution system for two coupled spatial metrics, we restricted ourselves to the simple case where the two metric tensors can be simultaneously diagonalized. Even this simplified case leads to very rich dynamics which can be conveniently described in terms of a mechanical model (represented in Fig.\ref{genplot}): two ``relativistic particles'', moving in a $(2+1)$-dimensional Lorentzian space, and connected by some non-linear ``spring'', {\textit{i.e.}} interacting via some bigravity potential ${\cal V}_{12}=(g_1g_2)^{1/4}V({\bf g}_1^{-1}{\bf g}_2)$.

One of the first important conclusion of our study is that the long-term behaviour of this coupled system crucially depends on the ability of the potential $V$ to confine the evolution of the relative lapse $\gamma=\log\left(N_2/N_1\right)$. Due to the former ``gauge nature''  of $\gamma$ (when the potential is absent), {\textit{i.e.}} the absence of kinetic terms for $\gamma$, the equation of motion of $\gamma$ is algebraic. We found that the continued existence (in the long term)  of a solution for $\gamma$ sensitively depends on the nature of the function $V(\gamma)$. For instance, we found that the potential $V({\bf g}_1^{-1}{\bf g}_2)$ derived from the five-dimensional brane constructions has only a marginal ability to confine the evolution of $\gamma$ to a limited range of variation. Indeed, we found that many (and maybe most) solutions of bi-cosmology, with such a brane potential, evolve, after some finite time, into a state where $\gamma$ quickly runs away towards infinity. By exploring the behaviour of physical observables near the moment where $|\gamma| \to \infty$, we have shown that the run away does not correspond to any observable singularity in either of the two metrics. We have argued that this run away only signals a breakdown of the effective four-dimensional description that we use. Indeed, it seems that, as $|\gamma| \to \infty$, some previously heavy modes become light and should now be taken into account in the effective action. 

Having understood the root of this run away behaviour, we have focused our physical study of bigravity on the class of potentials $V(\gamma)$ which have the strong-enough confining property with respect to $\gamma$. A simple example of a potential in this class is the ``quadratic plus quartic'' Pauli-Fierz-type potential (\ref{defpot}). Such a potential allows for solutions which evolve on long-time scales, without encountering any breakdown linked to a run away of $\gamma$. [Note, however, that we do not claim that this is true for all solutions. It is certainly possible to concoct initial data leading to a $\gamma$ runaway after a finite time.]
 
When using such $\gamma$-confining potentials (or when considering, as we do in most of the text, the effect of any potential up to times smaller than the moment of quick $\gamma$ runaway) we have found that the qualitative behaviour of generic cosmological solutions can be nicely understood in terms of the mechanical model of Fig.\ref{genplot}. For instance, when the separating vector $\delta^{\mu}=\beta^{\mu}-\alpha^{\mu}$ between the two ``particles'' is spacelike, and the potential is polynomial in $\delta^{\mu}$ (and attractive, as the modified Pauli-Fierz potential (\ref{defpot})), the coupled motion of $\alpha^{\mu}$ and $\beta^{\mu}$ is similar to slow-roll inflation. The separating vector $\delta^{\mu}$ plays the role of the inflaton, and drives an exponential-type expansion of the vertical position of the ``center of mass'' ${1 \over 2}\sigma={1 \over 2}(\beta+\alpha)$ (which represents the average volume of the two metrics). A qualitatively new feature of this type of bigravity slow-roll inflation is its growing anisotropy. As the connecting vector gets smaller, each metric tends to expand more and more differently in three spatial directions (linked to the ``direction'' of the vector $\delta^{\mu}$). This anisotropic slow-roll inflation ends up in a regime where the ``spring'' connecting the two ``particles'' makes them oscillate along a spacelike direction (see Fig.\ref{spaceplotnum}). Similarly to the oscillatory period following slow-roll (for a chaotic inflation type potential, {\textit{e.g.}} $V(\phi)={1 \over 2}m^2\phi^2$), these bigravity oscillations lead to a power-law expansion law. We expect such anisotropic accelerating solutions to exist in a general multigravity theory, and thus also in the brane-induced model   \cite{Dvali:2000hr,Dvali:2000xg}, which, as we said above, can be viewed as the $N \to \infty$ limit of the particular ``nearest neighbour'' interactions multigravity model.

When the separating vector $\delta^{\mu}=\beta^{\mu}-\alpha^{\mu}$ is timelike, we found (when using the $\gamma$-confining potential (\ref{defpot})) a remarkable phenomenon of ``locking'' of the two metrics. [We have indicated in Eqs.(\ref{dgcond}), (\ref{dcond}), (\ref{lockedrates}) the general conditions under which this ``locking'' phenomenon occurs.] In the visual language of Fig.{\ref{genplot}}, the two ``particles'' lock in a perpetual ``chasing'' configuration where their vertical separation tends to a non-zero constant, while their ``center of mass''  continues to move upwards. In bimetric language, the locking corresponds to a bi-de-Sitter configuration: each metric expands exponentially, and the two expansion rates are equal (in the average proper time). Contrary to the spacelike case where anisotropies played an important role, here this configuration is obtained for isotropically expanding metrics. It would be interesting to explore the basin of attraction of such a locked state among generic (timelike-type)  bigravity evolutions. As indicated above, these locked configurations admit (provided some system of $N-1$ equations for $N-1$ unknowns admit a real solution) a multi-de-Sitter generalization in a general multigravity model. [As we mentioned, this is a way of interpreting the solution of   \cite{Deffayet:2000uy,Deffayet:2001pu}.] Since we found that the bigravity locked solution was stable, we expect this feature to extend to the multigravity case.  
 
From the phenomenological point of view, one of the major conclusions of this work is that bigravity cosmologies generically exhibit a period of cosmic acceleration for one or both of the metrics. This conclusion applies even to the case of ``bad'' potentials which cannot permanently constrain the evolution of $\gamma$ to a bounded range. This result suggests that bigravity could be the origin of the observed cosmic acceleration, {\textit{i.e.}} that it could be the the source of {\textit{dark energy}}. In other words bigravity naturally defines a kind of ``tensor quintessence''. In brane models, the mass parameter $m^2$ appearing in the potential $V$ is an exponentially decreasing function of the interbrane distance. It is therefore not unnatural to have an $m^2$ as small as it is required to explain the observed cosmic acceleration ({\textit{i.e.}} $m \sim 10^{-33}$eV)\footnote{Let us note that the parameter $\mu$ appearing in (\ref{a1}) would then be $\mu \sim 10^{-3}$eV.}. 

Our preliminary studies of the transition between matter domination and vacuum domination seem to indicate that (at least for classes of potentials) this transition can be as smooth as in the usually considered dark-energy models (such as a cosmological constant, or some type of scalar quintessence). It is, however, interesting to note that, at least in the spacelike separated case, bigravity makes qualitatively new predictions: it predicts a growing anisotropy of the expansion of the universe. It would be interesting to study the imprint of this phenomenon (which started to take place only ``recently'', {\textit{i.e.}} for redshifts $z \lesssim 0.5$) on observable phenomena, and notably on the Cosmic Microwave Background.

Finally,   on a more speculative view, it would be interesting to explore the possibility that bigravity explains the primordial inflation needed to explain the gross features of our universe. For this, one would probably need a mass scale of order $m \sim 10^{-6} M_{\rm Pl}$. If we contemplate a ``spacelike'' scenario, the needed large initial value of $\delta^{\mu}$ to have a long stage of bigravity slow roll inflation, might be naturally provided by the recently discovered generic chaotic behaviour taking place in (bulk) string/M cosmology \cite{Damour:2000wm,Damour:2000hv,Damour:2001sa}. Indeed, the chaotic behaviour naturally leads, near $t \sim t_{\rm string}$ to very large ``oscillations'' in the (logarithmic) scale factors $\alpha^{\mu}$ of the metric (considered at some spatial points). When comparing two metrics (either at two different bulk points, or on two branes) it is then natural to reach large values of $\beta^{\mu}-\alpha^{\mu}$. On the other hand, if we contemplate a ``timelike'' scenario, the bi-de-Sitter locked configuration might naturally explain primordial inflation. In this case, one still needs an exit mechanism (which could be provided by some instability linked to $\gamma$ in the case where the potential $V(\gamma)$ cannot indefinitely succeed in confining $\gamma$).

\vskip1cm

\textbf{Acknowledgments:} I.K. would like to thank G.G. Ross for helpful discussions.  A.P. would like to thank R. Madden, H.P. Nilles, M. Peloso and  L. Pilo  for helpful discussions. I.K. is supported in part by PPARC rolling grant PPA/G/O/1998/00567 and  EC TMR grants HPRN-CT-2000-00152 and  HRRN-CT-2000-00148. A.P. acknowledges IHES for financial support under the Hodge Fellowship scheme.

\vskip1cm

\newpage

\def\theequation{A.\arabic{equation}}
\setcounter{equation}{0}
\vskip0.8cm
\noindent
{\Large \bf Appendix A: Equations of motion and constraints}
\vskip0.4cm
\noindent


The Hamiltonian of the system is:
\ba
{\mathcal H}=e^{\sigma/2}\left[{3 \over 4}\cosh\left({\gamma-\delta \over 2}\right)(-p_{\sigma}^2-p_{\delta}^2+p_{\Sigma_+}^2+p_{\Sigma_-}^2+p_{\Delta_+}^2+p_{\Delta_-}^2)e^{-\sigma}\right.\nonumber\\
\left.+{3 \over 2}\sinh\left({\gamma-\delta \over 2}\right)(-p_{\sigma}p_{\delta}+p_{\Sigma_+}p_{\Delta_+}+p_{\Sigma_-}p_{\Delta_-})e^{-\sigma}+V\right]
\ea
where the canonical momenta are defined as:
\ba
p_{\Delta_\pm}={2 \over 3}e^{\sigma/2}\left[\cosh\left({\gamma-\delta \over 2}\right)\dot{\Delta}_{\pm}-\sinh\left({\gamma-\delta \over 2}\right)\dot{\Sigma}_{\pm}\right]\\
p_{\delta}={2 \over 3}e^{\sigma/2}\left[\cosh\left({\gamma-\delta \over 2}\right)\dot{\delta}-\sinh\left({\gamma-\delta \over 2}\right)\dot{\sigma}\right]\\
p_{\sigma}={2 \over 3}e^{\sigma/2}\left[\cosh\left({\gamma-\delta \over 2}\right)\dot{\sigma}-\sinh\left({\gamma-\delta \over 2}\right)\dot{\delta}\right]
\ea

The equations of motion in the ``averaged proper time'' $t$ for the full Routhian are:

\begin{itemize}

\item $\delta$ equation

\be
{2 \over 3}e^{-\sigma/2}{d \over dt}\left[e^{\sigma/2}\left(\cosh\left({\gamma-\delta \over 2}\right)\dot{\delta}-\sinh \left({\gamma-\delta \over 2}\right)\dot{\sigma}\right)\right]={\de V \over \de \gamma}+{\de V \over \de \delta}
\ee

\item $\sigma$ equation

\be
{2 \over 3}e^{-\sigma/2}{d \over dt}\left[e^{\sigma/2}\left(\cosh\left({\gamma-\delta \over 2}\right)\dot{\sigma}-\sinh \left({\gamma-\delta \over 2}\right)\dot{\delta}\right)\right]=V
\ee

\item $\Delta_{\pm}$ equations

\be
e^{-\sigma/2}{d \over dt}\left[p_{\Sigma_{\pm}}\tanh\left({\gamma-\delta \over 2}\right)-{2e^{\sigma/2} \over 3\cosh\left({\gamma-\delta \over 2}\right)}\dot{\Delta}_{\pm}\right]={\de V \over \de \Delta_{\pm}}
\ee

\item $\gamma$ constraint

\ba
-{3 (p_{\Sigma_+}^2+p_{\Sigma_-}^2) \over 8}e^{-\sigma}{\tanh \left({\gamma-\delta \over 2}\right) \over \cosh \left({\gamma-\delta \over 2}\right) }+{e^{-\sigma/2} \over 2 \cosh^2 \left({\gamma-\delta \over 2}\right)}(p_{\Sigma_+}\dot{\Delta}_++p_{\Sigma_-}\dot{\Delta}_-)\nonumber\\+{1 \over 6}\sinh \left({\gamma-\delta \over 2}\right) \left[{\dot{\Delta}_+^2+\dot{\Delta}_-^2 \over \cosh^2\left({\gamma-\delta \over 2}\right)}+\dot{\sigma}^2+\dot{\delta}^2\right]-{1 \over 3}\cosh\left({\gamma-\delta \over 2}\right)\dot{\sigma}\dot{\delta}+{\de V \over \de \gamma}=0
\ea

\end{itemize}

In the above we have used the $\gamma$ constraint to simplify the $\delta$ equation of motion. We additionally have the Hamiltonian constraint ${\mathcal H}=0$:

\ba
{3 (p_{\Sigma_+}^2+p_{\Sigma_-}^2)e^{-\sigma}\over 4\cosh \left({\gamma-\delta \over 2}\right)}+{1 \over 3}\cosh \left({\gamma-\delta \over 2}\right) \left[{\dot{\Delta}_+^2+\dot{\Delta}_-^2 \over \cosh^2\left({\gamma-\delta \over 2}\right)}-\dot{\sigma}^2-\dot{\delta}^2\right]\nonumber\\+{2 \over 3}\sinh\left({\gamma-\delta \over 2}\right)\dot{\sigma}\dot{\delta}+V=0
\ea

The equations of motion in the ``averaged proper time'' $t$ for the case where $p_{\Sigma_+}=p_{\Sigma_-}=0$ are:

\begin{itemize}

\item $\delta$ equation

\be
{2 \over 3}e^{-\sigma/2}{d \over dt}\left[e^{\sigma/2}\left(\cosh\left({\gamma-\delta \over 2}\right)\dot{\delta}-\sinh \left({\gamma-\delta \over 2}\right)\dot{\sigma}\right)\right]={\de V \over \de \gamma}+{\de V \over \de \delta}
\ee

\item $\sigma$ equation

\be
{2 \over 3}e^{-\sigma/2}{d \over dt}\left[e^{\sigma/2}\left(\cosh\left({\gamma-\delta \over 2}\right)\dot{\sigma}-\sinh \left({\gamma-\delta \over 2}\right)\dot{\delta}\right)\right]=V
\ee

\item $r$ equation

\be
-{2 \over 3}e^{-\sigma/2}{d \over dt}\left({e^{\sigma/2}\dot{r} \over \cosh\left({\gamma-\delta \over 2}\right)}\right)=-{3 p_{\theta}^2\over 2 r^3}e^{-\sigma}\cosh \left({\gamma-\delta \over 2}\right)+{\de V \over \de r}
\ee

\item $\gamma$ constraint

\ba
{3 p_{\theta}^2\over 8 r^2}e^{-\sigma}\sinh \left({\gamma-\delta \over 2}\right)+{1 \over 6}\sinh \left({\gamma-\delta \over 2}\right) \left[{\dot{r}^2 \over \cosh^2\left({\gamma-\delta \over 2}\right)}+\dot{\sigma}^2+\dot{\delta}^2\right]\nonumber\\-{1 \over 3}\cosh\left({\gamma-\delta \over 2}\right)\dot{\sigma}\dot{\delta}+{\de V \over \de \gamma}=0
\ea

\end{itemize}

In the above we have used the $\gamma$ constraint to simplify the $\delta$ equation of motion. We additionally have the Hamiltonian constraint ${\mathcal H}=0$:

\ba
{3 p_{\theta}^2\over 4 r^2}e^{-\sigma}\cosh \left({\gamma-\delta \over 2}\right)+{1 \over 3}\cosh \left({\gamma-\delta \over 2}\right) \left[{\dot{r}^2 \over \cosh^2\left({\gamma-\delta \over 2}\right)}-\dot{\sigma}^2-\dot{\delta}^2\right]\nonumber\\+{2 \over 3}\sinh\left({\gamma-\delta \over 2}\right)\dot{\sigma}\dot{\delta}+V=0
\ea

Finally, the equations of motion in the ``averaged proper time'' $t$ for the case of $r=0$, when we have isotropic metrics, are:

\ba
{4 \over 3}~{d \over dt}\left[e^{\alpha+{\gamma \over 2}}\dot{\alpha}\right]&=&e^{\alpha+\beta \over 2}\left(V-{\de V \over \de \gamma}-{\de V \over \de \delta}\right)\\
{4 \over 3}~{d \over dt}\left[e^{\beta-{\gamma \over 2}}\dot{\beta}\right]&=&e^{\alpha+\beta \over 2}\left(V+{\de V \over \de \gamma}+{\de V \over \de \delta}\right)
\ea

We additionally have the two Friedmann constraints (which we have also used to simplify the above equations of motion):

\ba
{2 \over 3}\dot{\alpha}^2&=&e^{-{\gamma-\delta \over 2}}\left({V \over 2}-{\de V \over \de \gamma}\right)\\
{2 \over 3}\dot{\beta}^2&=&e^{{\gamma-\delta \over 2}}\left({V \over 2}+{\de V \over \de \gamma}\right)
\ea

\newpage

\def\theequation{B.\arabic{equation}}
\setcounter{equation}{0}
\vskip0.8cm
\noindent
{\Large \bf Appendix B: Perturbation analysis around the $b^{\mu}=0$ shift vector in the cosmological metric ansatz}
\vskip0.4cm
\noindent

In this Appendix we study under what conditions there might exist more solutions than the ``trivial'' solution $b^{\mu}=0$ discussed in the text. The eigenvalue problem:
\be
(g_1^{-1}g_2)^{M}_{N}e^{N}_A=\lambda_A e^{M}_A
\ee
considered for $b^{\mu} \neq 0$ but small,  can be related (to ${\cal O}(b^2)$) by standard techniques to the unperturbed  one (denoted by an overbar)  as:
\ba
\lambda_0&=&\bar{\lambda}_0\left(1-\sum_a {\bar{\lambda}_a (b^a)^2 \over \bar{\lambda}_0-\bar{\lambda}_a}\right)\\
\lambda_a&=&\bar{\lambda}_a\left(1+ {\bar{\lambda}_a (b^a)^2 \over \bar{\lambda}_0-\bar{\lambda}_a}\right) ~~~ ({\rm no~sum})
\ea
where $b^a=\bar{e}^a_{\mu}b^{\mu}$. The variation of the action when making a  perturbation is:
\ba
\delta S_m&=&-{1 \over 2}\sqrt{-g_2}T^{(2)~A}_{~~~A}~{\delta\lambda_A \over \lambda_A}\\
&=&\sqrt{-g_2}\sum_a  {\bar{\lambda}_a b^a \over \bar{\lambda}_0-\bar{\lambda}_a}(T^{(2)~0}_{~~~0}-T^{(2)~a}_{~~~a})\delta b^a
\ea
and hence:
\be
{\delta S_m \over \delta b^a}=\sqrt{-g_2} {\bar{\lambda}_a b^a \over \bar{\lambda}_0-\bar{\lambda}_a}(T^{(2)~0}_{~~~0}-T^{(2)~a}_{~~~a})~~~ ({\rm no~sum})\label{bvar}
\ee

As said in the text, the equation of motion of $b^{a}$ is the vanishing of (\ref{bvar}). There are only two possible ways to make the above quantity vanish: either $b^a=0$ or $T^{(2)~0}_{~~~0}-T^{(2)~a}_{~~~a}=0$. The first possibility means that $b^a=0$ is an isolated solution. Therefore, it is the second possibility which signals the threshold for the existence of new solutions, besides the trivial one. [We are assuming here that, as some parameters vary, the nonperturbative solutions can be made to coincide with the perturbative one.] On the other hand according to (\ref{energy1}),(\ref{energy2}):
\be
T^{(2)~0}_{~~~0}-T^{(2)~a}_{~~~a}=-2e^{-\sigma_1/4}(\de_{\mu_0}V-\de_{\mu_a}V)=-4e^{-\sigma_1/4}(\mu_0-\mu_a)\de_{\sigma_2}V
\ee
for potentials of the class $V=V(\sigma_1,\sigma_2)$. Therefore, a general necessary condition for the possible existence of non-perturbative solutions is that $\de_{\sigma_2}V = 0$ admits solutions.

\def\theequation{C.\arabic{equation}}
\setcounter{equation}{0}
\vskip0.8cm
\noindent
{\Large \bf Appendix C: Analytic solutions for the Pauli-Fierz potential in the timelike worldline separation limit}
\vskip0.4cm
\noindent

In this Appendix we sill study an analytic description of the solutions of the extreme timelike worldline separation  ({\textit{i.e.}} $r=0$) for the original Pauli-Fierz potential (\ref{PF}). As discussed in the text, the solutions exhibit an initial stage of acceleration for large $\delta$ and a period of deflation as $\delta \to 0$ for the first metric, while the second metric remains approximately flat.  We will split up the analysis of this system into the two above-mentioned asymptotic regions.

Let us note  the equations of state for the two metrics obtained by (\ref{state1}), (\ref{state2}):
\ba
w_1={P_1 \over \rho_1}=-1-2~{\delta-3\gamma \over \delta^2-6\delta+3\delta \gamma}\label{isostate1}\\
w_2={P_2 \over \rho_2}=-1+2~{\delta-3\gamma \over \delta^2+6\delta+3\delta \gamma}\label{isostate2}
\ea

\begin{itemize}

\item {\bf The cosmological initial singularity limit where   $\gamma \to -\infty$ and $\delta \to \infty$}

\end{itemize}

We have already qualitatively described in the text the behaviour of the system in this limit. Due to this motion in field space, we can see numerically that for all initial conditions, the first metric experiences accelerated expansion. As an example to illustrate the behaviour of the system in this limit, we will consider the case where $\gamma$ initially lies very near the $\gamma=-2-{1 \over 3}\delta$ line. Then, the solution follows this line in a very good approximation both back in time towards the initial singularity, as well as forward in time until $\delta \sim {\mathcal{O}}(1)$. From the $\beta$ Friedman equation  we get that $\dot{\beta}\approx 0$ and from the $\alpha$ constraint that: 
\be
\dot{\alpha}^2=-3m^2(\gamma+2)e^{-(2\gamma+3)}
\ee
But approximately $\dot{\alpha}=-\dot{\delta}=3\dot{\gamma}$, so we have a differential equation for $\gamma$. Integrating this we get:
\ba
{\rm Erf}(\sqrt{-(\gamma+2)})=1-{t\over t_{cr}}&\\
\Rightarrow~~\gamma=-2-\left[{\rm Erf^{(-1)}}\left(1-{t\over t_{cr}}\right)\right]^2 &~~\to~~\log\left(t \over t_{cr}\right) 
\ea
where $t_{cr}={1\over m}\sqrt{3\pi \over e}$, ${\rm Erf}(x)={2 \over \sqrt{\pi}}\int_0^x e^{-y^2}dy$ is the error function, ${\rm Erf^{(-1)}}(x)$ the inverse error function. The limit we have indicated is at $t \to 0$.  From the properties of the error function we have that if $t \to 0^+$ then $\gamma \to -\infty$. In the other limit that  $t \to t_{cr}$ we get that $\gamma \to -2$. However, we never reach the latter limit, since close to that, our approximation breaks down because $\delta \sim {\mathcal{O}}(1)$. In the region that this approximation is valid, we have:
\be
\delta=-3(\gamma+2)~~,~~\sigma=-\delta+2C~~,~~\alpha=-\delta+C~~,~~\beta=C
\ee
where $C$ is an integration constant. We see that the first metric expands, but one should describe this expansion in this metric's proper time. Asymptotically, for $t \to 0$ we have:
\be
dt_1=e^{-\gamma/2}dt~~\Rightarrow~~t_1=2\sqrt{t_{cr}t}
\ee
On the other hand the proper time of the second metric is:
\be
dt_2=e^{\gamma/2}dt~~\Rightarrow~~t_2={2 \over 3 \sqrt{t_{cr}}}t^{3/2}
\ee
Thus, the first metric is  intrinsically inflating with the scale factor behaving as  $e^{\alpha/3} \sim t_1^2$, while the second metric is approximately flat.

The equation of state (\ref{isostate1}), in  this limiting case which we are examining, is for the first metric:
\be
w_1=-1+{1 \over 3}~{\gamma+1 \over \gamma+2}~~\to~~-{2 \over 3}
\ee
while for the second one, (\ref{isostate2}) leads to  $w_2 \to -\infty$, since the  $\gamma=-2-{1 \over 3}\delta$ line is the root of the denominator.

Let us now check if the obtained solution is stable against perturbations of $r$. The variation of the Routhian in quadratic order is:
\be
\delta {\mathcal L}={1 \over 3}e^{\sigma /2}\left({\dot{r}^2 \over \cosh\left(\gamma-\delta \over 2\right)}-{1 \over 3}r^2\right)
\ee
From the extremization of this action we get the following motion for $r$:
\be
r=C_1J_0\left({1 \over 6}\sqrt{e \over 2\pi} m^2t^2\right)+C_2Y_0\left({1 \over 6}\sqrt{e \over 2\pi} m^2 t^2\right)
\ee
which is growing only logarithmically as $t \to 0$. Thus, the solution is stable in very good approximation. This is in agreement with our numerical study.

\begin{itemize}

\item {\bf The final state limit where $\gamma \to -\infty$, $\delta \to 0$}

\end{itemize}

The final state of the evolution of the system is independent of the initial conditions. As $\delta \to 0$ we find that  $\gamma \to -\infty$ and the first metric deflates and becomes asymptotically flat in infinite proper time, while the second experiences a finite (in proper time) period of inflation. The runaway of $\gamma$ in this limit is an unavoidable fact of the Pauli-Fierz potential, as we have discussed in the text, because the action ceases to have an extremum at finite $\gamma$, when one of the two worldline velocities tends to zero. In order to study this limit we need to do a different approximation to the equations of motion. Combining the two Friedman constraints (\ref{F1}), (\ref{F2}) and keeping leading terms we get the following relation:
\be
{\dot{\alpha}^2 \over \dot{\beta}^2}=e^{-\gamma}
\label{ratio}
\ee 
The equations of motion on the other hand are approximated by:
\ba
{4 \over 3}{d \over dt}\left(e^{\alpha+\gamma/2}\dot{\alpha}\right)={\gamma \over 3}e^{\alpha+\beta \over 2}\\
{4 \over 3}{d \over dt}\left(e^{\beta-\gamma/2}\dot{\beta}\right)=-{\gamma \over 3}e^{\alpha+\beta \over 2}
\ea
Using (\ref{ratio}) we have:
\ba
{d \over dt}\left(e^{\alpha}\dot{\beta}\right)={m^2 \over 2}e^{\alpha+\beta \over 2}\log{\dot{\beta} \over \dot{\alpha}}\\
{d \over dt}\left(e^{\beta}\dot{\alpha}\right)=-{m^2 \over 2}e^{\alpha+\beta \over 2}\log{\dot{\beta} \over \dot{\alpha}}
\ea
Then since $\dot{\alpha}\dot{\beta} \to 0$, we can neglect this term and obtain the system:
\ba
\ddot{\beta}={m^2 \over 2}e^{\beta-\alpha\over 2}\log{\dot{\beta} \over \dot{\alpha}} \approx {m^2 \over 2}\log{\dot{\beta} \over \dot{\alpha}}\\
\ddot{\alpha}=-{m^2 \over 2}e^{-{\beta-\alpha \over 2}}\log{\dot{\beta} \over \dot{\alpha}}\approx -{m^2 \over 2}\log{\dot{\beta} \over \dot{\alpha}}
\ea
From this system we get:
\be
\dot{\beta}=\dot{\alpha}\left({C_1 \over \dot{\alpha}}-1\right)
\ee
which putting back into the second equation gives:
\be
\ddot{\alpha}=-{m^2 \over 2}\log\left({C_1 \over \dot{\alpha}}-1\right)
\ee
This can be solved in the region where $\dot{\alpha} \to C_1$ and gives:
\be
\dot{\alpha}={C_1 \over 1+{\rm li}^{(-1)}\left({m^2  \over 2C_1}(t-t_0)\right)} \approx C_1 \left[1-{m^2  \over 2C_1}(t-t_0)\log \left({m^2  \over 2C_1}|t-t_0|\right)\right] 
\ee
where ${\rm li}(x)=\int_0^x {dy \over \log y}$ is the logarithmic integral and ${\rm li}^{(-1)}(x)$ its inverse function. The limits which have been used are that for $x \to 0^+$, ${\rm li}(x) \to {x \over \log x}$ and for $x \to 0^-$, ${\rm li}^{(-1)}(x) \to x \log |x|$. This shows that $\ddot{\alpha}$ diverges to $-\infty$. Asymptotically, the function  $\alpha$ is:
\be
\alpha \approx C_2+C_1 (t-t_0)-{m^2 \over 4}(t-t_0)^2\log \left({m^2  \over 2C_1}|t-t_0|\right)
\ee
The asymptotics for the $\beta$ function is:
\be
\beta=C_2+{m^2 \over 4}(t-t_0)^2\log \left({m^2  \over 2C_1}|t-t_0|\right)
\ee
and for $\delta$:
\be
\delta=-C_1 (t-t_0)
\ee
On the other hand $\gamma$ is:
\ba
\gamma=2\log \left({C_1 \over \dot{\alpha}}-1\right)~~\Rightarrow~~\gamma&=&2\log \left[{\rm li}^{(-1)}\left({m^2  \over 2C_1}(t-t_0)\right)\right]\nonumber\\&\approx& 2\log \left[{m^2  \over 2C_1}(t-t_0)\log \left({m^2  \over 2C_1}|t-t_0|\right)\right]
\ea

The proper time in the first metric is:
\be
dt_1=e^{-\gamma/2}dt~~\Rightarrow~~t_1 \sim {2C_1 \over m^2}\log \left|\log\left({m^2 \over 2C_1}|t-t_0|\right)\right|
\ee
thus, the singularity point $t_0$ in the proper time of the first metric is at $t_1 \to \infty$. On the other hand, the proper time for the second metric is:
\be
dt_2=e^{-\gamma/2}dt~~\Rightarrow~~t_2 = t_{20}  + {2C_1 \over m^2}\left[{m^2 \over 2C_1}(t-t_0)\right]^2 \log \left[{m^2 \over 2C_1}|t-t_0|\right]
\ee
and it is finite at $t \to t_0$.

Thus, the first metric is intrinsically deflating  with the scale factor behaving as:
\be
e^{\alpha/3} \sim \left(1-{2C_1^2 \over 3 m^2}e^{-e^{m^2t_1 \over 2C_1}}\right)
\ee
while the second is  intrinsically inflating as:
\be
e^{\beta/3} \sim e^{{C_1 \over 6}t_2}
\ee

The equation of state (\ref{isostate1}) on the first metric, in this limit, is:
\be
w_1=2~ {1 \over \delta} ~~\to~~+\infty 
\ee
while on the second one (\ref{isostate2}) leads to the opposite effect:
\be
w_2=-2~ {1 \over \delta} ~~\to~~-\infty 
\ee

\def\theequation{D.\arabic{equation}}
\setcounter{equation}{0}
\vskip0.8cm
\noindent
{\Large \bf Appendix D: The general $p \neq 0$ case evolution for the brane motivated potential}
\vskip0.4cm
\noindent

In this Appendix we will discuss the evolution of the system for the brane motivated potential for the general case where the ``initial kinetic energy'' $p$ is non-zero. The solution of the equations of motion is given by (\ref{x2sol}), (\ref{x1sol}). The initial ``incoming'' solutions are then for $\tau \ll 0$ where the kinetic energy of the system is large in comparison with the potential energy. Then  we can write the solutions of $\sigma$ and $r$ using (\ref{x1x2rs}) and separating the various constants (and subleading terms) as:  
\begin{itemize}
\item For $p>0$
\be
\sigma \to {3 \over 7}(3-\sqrt{2})p\tau + \cdots ~~~,~~~r \to \sigma +\cdots
\ee
\item For $p<0$
\be
\sigma \to -{3 \over 7}(3+\sqrt{2})p\tau + \cdots ~~~,~~~r \to -\sigma +\cdots
\ee
\end{itemize}
On the other hand, the final state solutions are  for $\tau \to 0$ when, due to the expansion of the two metrics, the kinetic energy of the system has become subdominant in comparison with the potential energy. Then  we have asymptotically for both signs of $p$:
\be
\sigma \to -{18 \over 7}\log |p\tau| + \cdots ~~~,~~~r \to -{\sqrt{2} \over 3}\sigma +\cdots
\ee
Note that the solutions in the latter epoch,  have the same scaling law as the ones for $p=0$. In the above language we see that in the case where $p>0$, $r$ initially increases until a maximum value and then shrinks to zero. On the other hand for $p < 0$ $r$ always increases. In all cases $\sigma$ increases and thus the volume of each metric expands.

Now, we need  to go back to proper time $t$ to see the behaviour of our solutions. For $\tau \ll 0$ both cases have $\sigma = c\tau+\cdots$ with different constants $c$, whose value does not have any significance as we will see in the following. Then the proper time and the expression of $\sigma$ as a function of it are:
\be
t = {2 \over c}~e^{c\tau+\cdots \over 2}~~~,~~~ \sigma=2 \log (p t) +\cdots
\label{pastasym}
\ee 
where again we ignored unimportant constants. By rescaling coordinates to absorb the latter constants we finally see that each metric has a {\textit{power-law behaviour}}, and if we parametrise it in the standard way as:
\be
ds^2=-dt^2+\sum_{\mu=1}^3(p t)^{2 p^{\mu}}(dx^{\mu})^2
\label{Kas1}
\ee
we have that the Kasner exponents $p^{\mu}$ for $p>0$ are:
\be
\renewcommand{\arraystretch}{1.5}
p^{\mu}_{(1)}=\left[\begin{array}{ccc}{1 \over 3}+{1 \over \sqrt{3}}\\{1 \over 3}-{1 \over \sqrt{3}}\\{1 \over 3}\end{array}\right]~~~{\rm and}~~~p^{\mu}_{(2)}=\left[\begin{array}{ccc}{1 \over 3}-{1 \over \sqrt{3}}\\{1 \over 3}+{1 \over \sqrt{3}}\\{1 \over 3}\end{array}\right]
\ee
and for $p<0$ they are the same with a flip on the sign of the second addendum of the first two exponents.  The above exponents satisfy the usual quadratic Kasner relation $\sum_{\mu} (p^{\mu})^2-\left(\sum_{\mu} p^{\mu}\right)^2=0$, as well as  $\sum_{\mu} p^{\mu}=1$, which means that each metric's volume $v_i$ expands as a function of its respective proper time as $v \propto t_i$, with $i=1,2$.  Thus, this evolution is highly anisotropic with:
\be
A_1=A_2=\sqrt{2}
\ee

One can notice at this point a potential paradox because an exact Kasner metric is known to be ``on the light cone'', {\textit{i.e.}} to  have $\dot{\alpha}^{\mu}\dot{\alpha}_{\mu}={1 \over 6}(\dot{\sigma}^2-\dot{r}^2)=0$. On the other hand, from the Hamiltonian  constraint (\ref{eq_H}) the same quantity should be very large for $r \gg 1$. This can be understood if we include next to leading order terms in our asymptotic solution. These will modify the $\sigma$ asymptotic (\ref{pastasym}) by a term linear in $t$ and also the $r$ asymptotic with a term linear in $t$ with different coefficient. This will immediately render  $\dot{\sigma}^2-\dot{r}^2$ very large as expected. 

For the case where $\tau \to 0$, we can express the proper time and $\sigma$ as a function of it as:
\be
t=C \tau^{-{2 \over 7}}~~~,~~~ \sigma=9 \log (p t) +\cdots
\ee
Thus, the exponents this time are:
\be
\renewcommand{\arraystretch}{1.5}
p^{\mu}_{(1)}=\left[\begin{array}{ccc}{3 \over 2}-\sqrt{{3 \over 2}}\\{3 \over 2}+\sqrt{{3 \over 2}}\\{3 \over 2}\end{array}\right]~~~{\rm and}~~~p^{\mu}_{(2)}=\left[\begin{array}{ccc}{3 \over 2}+\sqrt{{3 \over 2}}\\{3 \over 2}-\sqrt{{3 \over 2}}\\{3 \over 2}\end{array}\right]
\ee
The above exponents do not satisfy the quadratic (zero-mass-shell) Kasner relation, since  $\sum_{\mu} (p^{\mu})^2-\left(\sum_{\mu} p^{\mu}\right)^2=-{21 \over 2}$. In addition, $\sum_{\mu} p^{\mu}={9 \over 2}$, which means that each metric's volume $v_i$ expands as a function of its respective proper time as $v \propto t_1^{9/2}$, with $i=1,2$. Note that the latter volume expansion has an accelerating behaviour and is exactly the same as in the $p=0$ case.   The evolution is still anisotropic but slightly less than the Kasner case:
\be
A_1=A_2={2 \over 3}
\ee

Let us now check, as we did for the $p=0$ case,  if the obtained solution is stable against perturbations of $\delta$. We have to recalculate the quantities $A$ and $B$ of Eq.(\ref{ABdef})  and their limits for  $r \gg 1$. For $\tau \ll 0$ both signs of $p$ have:
\be
A=0 ~~~,~~~B\to {m^4 t^2 \over 12}e^{\sigma/2}
\ee
and hence we have $\delta=0$ and absolutely stable motion. On the other hand, for the case $\tau \to 0$, since the evolution is exactly the same as for the $p=0$ case, the solution is unstable at late times as described in the main text. This instability is again linked to a run away of $\gamma$ towards large values.

\end{document}